\documentstyle[epsfig,twocolumn,subfigure]{mn}

\title[The Sersic Virial Hyperplane]{The Sersic Virial Hyperplane}
\author[S. Capozziello et al.]{S.Capozziello$^{1}$, V. F. Cardone$^{2}$\thanks{Corresponding
author\,: {\tt winny@sa.infn.it}.}, R. Molinaro$^{2}$, V. Salzano$^1$ \\
$^1$ Dipartimento di Scienze Fisiche, Universit\`{a} degli Studi
di Napoli ``Federico II'' and INFN, Sezione di Napoli, \\
Complesso Universitario di Monte S. Angelo, Via Cinthia, Edificio
N, 80126 Napoli, Italy\\ $^2$ Dipartimento di Fisica ``E.R.
Caianiello'', Universit{\`{a}} di Salerno, Via S. Allende, 84081 -
Baronissi (Salerno), Italy}

\date{Accepted xxx, Received yyy, in original form zzz}

\begin{document}

\maketitle

\begin{abstract}

Early\,-\,type galaxies (ETGs) are known to possess a number of
quite useful scaling relations among their photometric and/or
kinematical quantities. We propose a unified picture reducing both
the fundamental plane and the photometric plane to suitable
projections of a single relation. Modelling the ETG as a two
component system, made out of a luminous Sersic profile and a NFW
dark halo, and applying the virial theorem, we are able to express
the velocity dispersion $\sigma_0$ as a function of the effective
intensity $\langle I_e \rangle$, the effective radius $R_e$ and
the Sersic index $n$. In a log\,-\,log plot, this relation reduces
to a hyperplane (i.e., a plane in a four dimensional configuration
space) which we dubbed the {\it Sersic Virial Hyperplane} (SVH).
The tilt of the SVH can be fully explained in terms of a
power\,-\,law scaling of the stellar (rather than the global)
mass\,-\,to\,-\,light ratio $\Upsilon_{\star}$ with the total
luminosity $L_T$, while the scatter is determined by those on the
$c\,-\,M_{vir}$ relation between the concentration $c$ and the
virial mass $M_{vir}$ of the dark halo component. In order to test
whether such the observed SVH is consistent with the theoretical
assumptions, we perform a detailed simulation of a large ETGs
sample reproducing the same photometric properties of a SDSS low
redshift ETGs catalog. It turns out that the simulated SVH is
fully consistent with the observed one thus validating our
theoretical approach.

\end{abstract}


\section{Introduction}

Elliptical and S0 galaxies (hereafter, collectively referred to as
early\,-\,type galaxies, ETGs) present a striking regularity in
their luminosity distribution. The ETG surface brightness is well
fitted by the well known $r^{1/4}$ profile \cite{deV48}, while
considerable better results are obtained using the Sersic
$r^{1/n}$ law \cite{Sersic}. From the photometric point of view,
therefore, ETG may be considered as characterized by only three
parameters, namely the slope $n$ of the Sersic profile, the
effective radius $R_e$ containing half of the total luminosity,
and the effective surface brightness $\mu_e$ defined as $\mu_e =
\mu(R_e)$, or equivalently the average surface intensity $\langle
I_e \rangle = (L_T/2)/(\pi R_e^2)$.

The ETG kinematic may be schematically characterized through its
central velocity dispersion $\sigma_0$ which, under suitable
assumptions on the mass profile, gives information on the mass and
hence the mass\,-\,to\,-\,light (hereafter $M/L$) ratio. Being a
mass tracer, it is reasonable to expect that $\sigma_0$ is
somewhat correlated to the total luminosity $L_T$, even if it is
difficult to forecast an analytic form for such a correlation,
given the subtleties of the luminous and dark components
modelling. It is therefore not surprising that empirical searches
for such a correlation were early undertaken. A remarkable success
was represented by the Faber\,-\,Jackson (FJ) $L_T \propto
\sigma_0^4$ relation \cite{FJ76}. The large scatter in the FJ law
led to the need for higher dimensional representations of ETGs.
Considering $n = 4$, there remains just three parameters
describing an ETG so that one could wonder whether a relation
exist among the photometric quantities $(R_e, \langle I_e
\rangle)$ and the kinematical $\sigma_0$. This relation were
indeed found \cite{DD87,7S87} and, when expressed in a logarithmic
scale, is just a plane soon dubbed the {\it fundamental plane}
(FP). It is worth noting that such a plane was not unexpected.
Indeed, a simple application of the virial theorem gives $R_e
\propto \sigma_0^2 \langle I_e \rangle^{-1}$, under the hypotheses
of constant $M/L$ ratio and structural homology. The observed FP
plane is however tilted, i.e. one indeed finds $R_e \propto
\sigma_0^{\alpha} \langle I_e \rangle^{-\beta}$, but with
$(\alpha, \beta) = (1.49, 0.75)$ rather than $(2, 1)$ as
forecasted before \cite{BernFP}. Such a tilt may be easily
explained introducing a power\,-\,law correlation $M/L \propto
L_T^{\gamma}$ (with $\gamma \simeq 0.27$), but interpreting the
origin of such a relation is a difficult and ambiguous task due to
proposals ranging from non\,-\,homology \cite{PS97}, to stellar
populations effects \cite{7S87}, from systematic variations in
kinematical structure \cite{vdM91,BBF92,Bus97} to a combination of
different terms \cite{TBB04}.

Although the de Vaucouleurs profile is a satisfactory fit, it is
well known that the Sersic profile has to be preferred
\cite{CCD93,GC97,PS97}. As such, forcing $n = 4$ in the fit may
systematically bias the estimate of $R_e$ and $\langle I_e
\rangle$ and hence affect the FP. Introducing $n$ increases the
number of parameters needed to describe ETGs leading to wonder
whether scaling relations exist. Actually, given the observational
difficulties in measuring $\sigma_0$, it is worth looking for
empirical correlations involving only the photometric parameters.
Interesting examples are the Kormendy relation (KR) between $R_e$
and $\mu_e$ \cite{K77} and the scalelength\,-\,shape relation
between $R_e$ and $n$ \cite{YC94}. However, just as the FJ
relation is a projection of the FP, both the $R_e$\,-\,$\mu_e$ and
the $R_e$\,-\,$n$ relations may be seen as projections of a more
fundamental law among these three photometric parameters. In
logarithmic units, such a relation indeed exists and it is a plane
referred to as the {\it photometric plane} (PhP) recently detected
in both near infrared \cite{KWKM00} and optical \cite{G02}. While
observationally the PhP is confirmed also at intermediate redshift
(La Barbera et al. 2004, 2005), a definitive theoretical
interpretation is still lacking. Modelling the stars in ETG as a
self\,-\,gravitating gas, Lima Neto et al. (1999) have recovered a
PhP like relation (referred to as the {\it entropic plane}) by
assuming that the specific entropy (i.e., the entropy for mass
unit) is constant for all ETGs. Later, M\'arquez et al. (2001)
derived an {\it energy\,-\,entropy} (or {\it mass\,-\,entropy})
line giving a possible explanation for the structural relations
among photometric parameters. Moreover, they also find out that
the specific entropy increases as a consequence of merging
processes so offering a possible way to test the model against the
observed variation of the PhP with redshift.

There are two general lessons to draw from the above short
summary. First, two dimensional scaling relations turn out to be
projections of a more general three dimensional law. It is
therefore worth wondering whether this also holds for the FP and
PhP being possible projections of a four parameters law. It is
worth noting that a first step towards this direction has been
attempted by Graham (2002) fitting a hyperplane (in logarithmic
units) to the set of parameters $(n, R_e, \langle I_e \rangle,
\sigma_0)$, but it was never prosecuted. On the other hand, from a
theoretical point of view, both the entropic plane and the FP are
tentatively explained on the implicit assumption that ETGs are in
a state of dynamical equilibrium so that the virial theorem
applies and the Boltzmann\,-\,Gibbs entropy may be evaluated.
Motivated by these considerations, here we investigate whether a
four dimensional relation among photometric and kinematic
quantities may come out as consequence of the virial theorem and
some assumptions on the stellar $M/L$ consistent with stellar
population synthesis models. Should such a relation exist, one
could thus reconcile both the FP and the PhP under the same
theoretical standard thus representing a valid tool to investigate
ETG formation theories.

As a first key ingredient, we describe in Sect. 2 the assumed mass
models for both the luminous and the dark components of ETGs. In
Sect. 3, we first compute the quantities entering the virial
theorem according to the model detailed above and then obtain the
four dimensional scaling relation we are looking for. In order to
test whether our derivation is consistent with what is observed,
we simulate ETG samples that are as similar as possible to a large
SDSS based sample following the procedure extensively discussed in
Sect. 4. The results of such a testing procedure are described in
Sect. 5, while Sect. 6 is devoted to fit our theoretical relation
to the observed data using a Bayesian approach. Discussion and
conclusions are finally held in Sect. 7.

\section{Modelling elliptical galaxies}

Although some recent results show the presence of thin discs in
inner regions of elliptical galaxies\footnote{S0 galaxies contain,
by definition, a thin disc, so that, strictly speaking, our
following discussion applies only to the bulge component. However,
neglecting this disc does not introduce any significant systematic
error. Moreover, our sample is mainly dominated by elliptical
galaxies so that we confidently neglect the disc component in S0
systems.}, ETGs may be well described as a luminous stellar
distribution embedded in a dark matter halo dominating the outer
mass profile. In the following, we will assume spherical symmetry
for both these components. While this is an acceptable hypothesis
for the halo, it is clearly an oversimplification for the
elliptical luminous component. Nevertheless, this will allow us to
get analytical expressions for the main quantities we are
interested in without dramatically affecting the main results.

\subsection{The Sersic profile}

Under the hypothesis of constant $M/L$, the surface density of the
luminous component may be easily obtained as\,:

\begin{equation}
\Sigma(R) = \Upsilon_{\star} I(R)
\label{eq: sigmalum}
\end{equation}
with $\Upsilon_{\star}$ the $M/L$ ratio and $I(R)$ the surface
luminosity density. As well known, the Sersic $r^{1/n}$ law
\cite{Sersic} is best suited to describe the surface brightness
distribution of elliptical galaxies \cite{CCD93,GC97,PS97}.
Motivated by these evidences, we will therefore set\,:

\begin{equation}
I(R) = I_e \exp{\left \{ -b_n \left [ \left ( \frac{R}{R_e} \right
)^{1/n} - 1 \right ] \right \}} \label{eq: ir}
\end{equation}
with $I_e$ the luminosity density at the effective radius $R_e$
and $b_n$ a constant defined such that the luminosity within $R_e$
is half the total luminosity. It is possible to show that $b_n$
may be found by solving \cite{C91}\,:

\begin{equation}
\Gamma(2n, b_n) = \Gamma(2n)/2 \label{eq: bn}
\end{equation}
where $\Gamma(a, z)$ is the incomplete $\Gamma$ function and
$\Gamma(a)$ the actual $\Gamma$ function. Useful approximating
formulae may be found in Graham \& Driver (2005) and references
therein, but we will exactly solve Eq.(\ref{eq: bn}) in the
following.

Assuming cylindrical symmetry, the luminosity profile within $R$
is\,:

\begin{equation}
L(R) = 2 \pi \int_0^R{I(R')R' dR'} = L_T \times \frac{\gamma(2n,
b_n x)}{\Gamma(2n)} \label{eq: lr}
\end{equation}
with $x \equiv R/R_e$ and

\begin{equation}
L_T = 2 \pi n I_e R_e^2 b_n^{-2n} {\rm e}^{b_n} \Gamma(2n)
\label{eq: lt}
\end{equation}
the total luminosity. The volume luminosity density $j(s)$ may be
easily obtained deprojecting $I(R)$. Defining $s = r/R_e$ (with
$r$ the radius in spherical coordinates), we get \cite{MC02}\,:

\begin{displaymath}
\nu(s) = - \frac{1}{\pi} \int_{s}^{\infty}{\frac{di}{dx}
\frac{dx}{\sqrt{x^2 - s^2}}}
\end{displaymath}
with $i(x) = I(R)/I_e$ and $\nu(s) = (R_e/I_e) j(s)$. Some algebra
finally leads to the following expressions for the mass density
$\rho_{\star}(s)$ and the mass profile $M_{\star}(s)$\,:

\begin{equation}
\rho_{\star}(s) = \frac{M_{\star}^T}{4 \pi R_e^3} \times
\frac{{\cal{I}}_\nu(s)}{{\cal{I}}_M(s)} \ , \label{eq: rhostar}
\end{equation}

\begin{equation}
M_{\star}(s) = M_{\star}^T \times
\frac{{\cal{I}}_M(s)}{{\cal{I}}_M(s)} \ , \label{eq: massstar}
\end{equation}
where we have denoted with $M_{\star}^T$ the total stellar mass\,:

\begin{equation}
M_{\star}^T = 4 (b_n/n) \Upsilon_{\star} I_e R_e^2
{\cal{I}}_M(\infty) \ , \label{eq: starmass}
\end{equation}
having used the abuse of notation

\begin{displaymath}
f(\infty) = \lim_{y \rightarrow \infty}{f(y)} \ .
\end{displaymath}
Finally, to get Eqs.(\ref{eq: rhostar}) and (\ref{eq: massstar}),
we have used the auxiliary functions\,:

\begin{equation}
{\cal{I}}_{\nu}(s) = \int_{s}^{\infty}{\frac{x^{(1 - n)/n}
\exp{\left [ -b_n \left ( x^{1/n} - 1 \right ) \right ]}}{\left
(x^2 - s^2 \right )^{1/2}} dx} \ , \label{eq: definu}
\end{equation}

\begin{equation}
{\cal{I}}_M(s) = \int_{0}^{s}{{\cal{I}}_{\nu}(s') s'^2 ds'} \ .
\label{eq: defim}
\end{equation}
Both these functions cannot be analytically expressed, but are
straightforward to be numerically evaluated.

\subsection{The dark halo}

Most of the kinematical tracers of the total gravitational
potential are usually affected by a severe degeneracy between the
luminous component and the dark one so that there are different
dark halo models able to fit the same data for a given stellar
mass profile. It is therefore important to rely on a physical
theory of halo formation to select models which are both
compatible with the data and also physically well motivated. From
this point of view, numerical simulations of galaxy formation in
hierarchical CDM scenarios are very helpful since they predict the
initial shape of the dark matter distribution. Here, we assume a
NFW profile \cite{NFW97} as initial dark matter halo and neglect
the effect of the baryons gravitational collapse. The main
features of the NFW model are\,:

\begin{equation}
\rho_{DM}(r) \equiv \frac{\rho_s}{x \ (1+x)^2} \ , \ x = r/r_s
\label{eq: nfwhalo}
\end{equation}

\begin{equation}
M_{DM}(r) = 4 \pi \rho_s r_{s}^{3} \ f(x) = M_{vir} f(x)/f(c) \ ,
\label{eq: nfwmass}
\end{equation}

\begin{equation}
f(x) \equiv \ln{(1+x)} - \frac{x}{1+x} \label{eq: deffnfw}\ ,
\end{equation}

\begin{equation}
c \equiv r_{vir}/r_s \ , \label{eq: rvir}
\end{equation}

\begin{equation}
M_{vir} = \frac{4 \pi \delta_{th}}{3} \rho_{crit} r_{vir}^3 \ ,
\end{equation}
where $c$ is the concentration parameter, $M_{vir}$ the virial
mass and $r_{vir}$ the virial radius\footnote{The virial radius is
defined such that the mean density within $r_{vir}$ is
$\delta_{th}$ times the critical density $\rho_{crit}$. According
to the concordance $\Lambda$CDM model, we assume a flat universe
with $(\Omega_m, \Omega_{\Lambda}, h) = (0.3, 0.7, 0.72)$ where
$\delta_{th} = 337$.}. The model is fully described by two
independent parameters, which we assume to be $c$ and $M_{vir}$.
Numerical simulations, supported by observational data, motivate a
correlation between $c$ and $M_{vir}$ so that the NFW model may be
considered as a single parameter family. Following Bullock et al.
(2001), we adopt\,:

\begin{equation}
c = 15 - 3.3 \log{\left (M_{vir}{10^{12} h^{-1} \ {\rm M_{\odot}}}
\right )} \label{eq: cmvir}
\end{equation}
with a log normal scatter $\delta \log{c} \simeq 0.11$.

The NFW model is not the only model proposed to fit the results of
numerical simulations. Some authors \cite{Moore98,Ghigna00} have
proposed models with a central slope steeper than the NFW one. On
the other hand, it is also possible that the inner slope does not
reach any asymptotic value with the logarithmic slope being a
power\,-\,law function of the $r$ \cite{N04,PoLLS} or that the
deprojected Sersic profile also fits the numerical dark matter
haloes \cite{Merr05,G06a,G06b}. However, the difference between
all these models and the NFW one is very small for radii larger
than 0.5$\%$\,-\,1$\%$ the virial radius so that we will not
consider models other than the NFW one.

\section{The virial theorem}

Elliptical galaxies are known to be characterized by scaling
relations among their kinematic and structural parameters. In an
attempt to investigate whether such empirical laws may be
recovered under a single theoretical scheme, we can rely on the
hypothesis of statistical equilibrium. In such an assumption, the
virial theorem holds\,:

\begin{equation}
2 K + W = 0 \label{eq: virtheo}
\end{equation}
with $K$ and $W$ the total kinetic and potential energy
respectively. Both these quantities may be evaluated given the
assumed spherical symmetry as we detail in the following.

\subsection{Kinetic energy}

Neglecting the net rotation velocity of the system (which is
reasonable, given the low values of $v_c$ in elliptical galaxies),
the total kinetic energy is given as \cite{BT87}\,:

\begin{equation}
K = 3 \pi \int_{0}^{\infty}{\Sigma(R) \sigma_p^2(R) R dR}
\label{eq: defk}
\end{equation}
with $\Sigma(R)$ the star surface density and $\sigma_p(R)$ the
luminosity weighted velocity dispersion projected along the line
of sight. For a spherically symmetric system, assuming an
isotropic velocity dispersion tensor, it is\,:

\begin{equation}
\sigma_p^2(R) = \frac{2}{\Upsilon_{\star} I(R)}
\int_{R}^{\infty}{\frac{\rho_{\star}(r) G M_{tot}(r) \sqrt{r^2 -
R^2}}{r^2} dr} \ , \label{eq: defsigmap}
\end{equation}
with $M_{tot}(r) = M_{\star}(r) + M_{DM}(r)$ the total mass
profile. As a first step, it is convenient to split the integral
in Eq.(\ref{eq: defsigmap}) in two terms separating the dark halo
contribution from the luminous one. It is then easy to get\,:

\begin{eqnarray}
\int_{R}^{\infty}{\frac{\rho_{\star}(r) G M_{\star}(r) \sqrt{r^2 -
R^2}}{r^2} dr} & = &  \frac{\left ( M_{\star}^{T} \right )^2}{4
\pi R_e^3
{\cal{I}}_M^2(\infty)} \nonumber \\
~ & \times &  \int_{R}^{\infty}{\frac{{\cal{I}}_{\nu}(s)
{\cal{I}}_M(s)}{s^2 \left (s^2 - x^2 \right )^{-1/2}} ds} \ ,
\nonumber
\end{eqnarray}

\begin{eqnarray}
\int_{R}^{\infty}{\frac{\rho_{\star}(r) G M_{DM}(r) \sqrt{r^2 -
R^2}}{r^2} dr} & = & \frac{M_{\star}^{T} M_{vir}}{4 \pi R_e^3
{\cal{I}}_M(\infty) f(c)} \nonumber \\
~ & \times & \int_{R}^{\infty}{\frac{{\cal{I}}_{\nu}(s) f(s,
R_e/r_s)}{s^2 \left ( s^2 - x^2 \right )^{-1/2}} ds} \ , \nonumber
\end{eqnarray}
where $f(s, R_e/r_s)$ is obtained by replacing $x$ with $(R_e/r_s)
s$ into Eq.(\ref{eq: deffnfw}). Adding the two terms above and
using Eqs.(\ref{eq: ir}) and (\ref{eq: starmass}) , we can finally
write\,:

\begin{eqnarray}
\sigma_p^2(R) & = & \frac{2 b_n}{n \pi} \frac{G M_{\star}^T}{R_e}
\\ ~ & \times & \frac{{\cal{I}}_{\sigma}^{\star}(x, n) + \left
(M_{vir}/M_{\star}^T \right ) f^{-1}(c) {\cal{I}}_{\sigma}^{DM}(x,
n, R_e/r_s)}{\exp{\left [ - b_n \left (x^{1/n} - 1 \right ) \right
]}} \nonumber \label{eq: sigmap}
\end{eqnarray}
with $x = R/R_e$ and\,:

\begin{displaymath}
{\cal{I}}_{\sigma}^{\star}(x, n) \equiv
\int_{x}^{\infty}{\frac{{\cal{I}}_{\nu}(s) {\cal{I}}_M(s)}{s^2
\left (s^2 - x^2 \right )^{-1/2}} ds} \ ,
\end{displaymath}

\begin{displaymath}
{\cal{I}}_{\sigma}^{DM}(x, n, R_e/r_s) \equiv
\int_{x}^{\infty}{\frac{{\cal{I}}_{\nu}(s) f(s, R_e/r_s)}{s^2
\left (s^2 - x^2 \right )^{-1/2}} ds} \ .
\end{displaymath}
As we will see later, $\sigma_p^2(R)$ does not enter the
applications we are interested in. It is rather its central value
$\sigma_0^2$ averaged within a circular aperture of radius $R_e/8$
which is measured by the galaxy spectrum. We therefore evaluate\,:

\begin{equation}
\sigma_0^2 = \frac{\int_{0}^{R_e/8}{\rho_{\star}(r) M_{tot}(r)
\left [ {\cal{F}}_0(R_e/8r) - {\cal{F}}_0(0) \right ] dr}}{2
\int_{0}^{R_e/8}{\rho_{\star}(r) \left [ {\cal{A}}_0(R_e/8r) -
{\cal{A}}_0(0) \right ] r dr} } \ , \label{eq: sz}
\end{equation}
with \cite{DK03}\,:

\begin{displaymath}
{\cal{F}}_0(y) = \left \{
\begin{array}{ll}
y \sqrt{1 - y^2} + \arcsin{y} - \pi/2 & y < 1 \\
0 & y > 1
\end{array}
\right . \ ,
\end{displaymath}

\begin{displaymath}
{\cal{A}}_0(y) = \left \{
\begin{array}{ll}
\arcsin{y} & y < 1 \\
\pi/2 & y > 1
\end{array}
\right . \ .
\end{displaymath}
When evaluating Eq.(\ref{eq: sz}), we make the simplifying
assumptions $M_{tot} \simeq M_{\star}$ since the dark halo mass is
negligible with respect to the stellar one over the range
interested by the integral. By such an assumption, it is easy to
get\,:

\begin{equation}
\sigma_0^2 = \frac{G M_{\star}^T}{R_e} \times {\cal{I}}_0(n)
\label{eq: sigmazero}
\end{equation}
having defined\,:

\begin{equation}
{\cal{I}}_0(n) = \frac{1}{{\cal{I}}_M(\infty)}
\frac{\int_{0}^{1/8}{{\cal{I}}_{\nu}(s) {\cal{I}}_{M}(s) \left [
{\cal{F}}_0(1/8s) - {\cal{F}}_0(0) \right ] ds}}{2
\int_{0}^{R_e/8}{{\cal{I}}_{\nu}(s) \left [ {\cal{A}}_0(1/8s) -
{\cal{A}}_0(0) \right ] s ds}} \ . \label{eq: defiz}
\end{equation}
As a next step, we insert Eq.(\ref{eq: sigmap}) into Eq.(\ref{eq:
defk}) and then split the integral in two terms, the first one
originated by the product $\Sigma(s) \times
{\cal{I}}_{\sigma}^{\star}(s)$, and the second one due to
$\Sigma(s) \times {\cal{I}}_{\sigma}^{DM}(s)$. By using
Eq.(\ref{eq: sigmazero}), we finally obtain\,:

\begin{equation}
K = \frac{1}{2} M_{\star}^T \sigma_0^2 \times k({\bf p})
\label{eq: endk}
\end{equation}
with ${\bf p}$ denoting the set of parameters

\begin{displaymath}
{\bf p} = (n, R_e, I_e, \Upsilon_{\star}, M_{vir}, c) \ ,
\end{displaymath}
where we have set\,:

\begin{equation}
k({\bf p}) = \frac{3}{2} \left [ k_{\star}(n, R_e, I_e) +
\frac{M_{vir}/M_{\star}^T}{f(c)} k_{\star}^{DM}(n, R_e/r_s) \right
] \ , \label{eq: defkpar}
\end{equation}

\begin{equation}
k_{\star}(n) = \frac{1}{{\cal{I}}_0(n) {\cal{I}}_M(\infty)}
\int_{0}^{\infty}{{\cal{I}}_{\sigma}^{\star}(x, n) x dx} \ ,
\label{eq: defkstar}
\end{equation}

\begin{equation}
k_{\star}^{DM}(n, R_e/r_s) = \frac{1}{{\cal{I}}_0(n)
{\cal{I}}_M(\infty)} \int_{0}^{\infty}{{\cal{I}}_{\sigma}^{DM}(x,
R_e/r_s) x dx} \ . \label{eq: defkstardm}
\end{equation}
Note that, although $I_e$ and $\Upsilon_{\star}$ do not explicitly
enter the above equations, they are however included as parameters
since they determine the total stellar mass $M_{\star}^T$ because
of Eq.(\ref{eq: starmass}). On the other hand, $r_s$ is not
counted as an independent parameter since it is determined as
function of $M_{vir}$ and $c$. Should we use Eq.(\ref{eq: cmvir}),
the virial mass $M_{vir}$ would be the only parameter related to
the dark halo properties.

\subsection{The gravitational energy}

The computation of $W$ may be carried out in a similar way
starting from the definition \cite{BT87}\,:

\begin{equation}
W = - 4 \pi G \int_{0}^{\infty}{\rho_{tot}(r) M_{tot}(r) r dr} \ .
\label{eq: defw}
\end{equation}
Splitting the total density and mass as the sum of the luminous
and dark components, after some algebra, one gets\,:

\begin{equation}
W = - \frac{G M_{\star}^{T 2}}{R_e^2} \times w({\bf p}) \label{eq:
endw}
\end{equation}
with ${\bf p}$ denoting the same set of parameters as above. The
dimensionless quantity $w({\bf p})$ is defined as\,:

\begin{eqnarray}
w({\bf p}) & = & w_{\star}(n) \nonumber \\
~ & + & \left ( \frac{M_{vir}}{M_{\star}^T} \right )^2 \left (
\frac{R_e}{r_{vir}} \right ) w_{DM}(c) \nonumber \\
~ & + & \left ( \frac{M_{vir}}{M_{\star}^T} \right )
w_{\star}^{DM}(c, n, R_e/r_s)  \\
~ & + & \left ( \frac{M_{vir}}{M_{\star}^T} \right )
w_{DM}^{\star}(c, n, R_e/r_s) \ . \nonumber \label{eq: defwpar}
\end{eqnarray}
It is then only a matter of algebra to demonstrate that\,:

\begin{equation}
w_{\star}(n) = \frac{1}{{\cal{I}}_{M}^2(\infty)}
\int_{0}^{\infty}{{\cal{I}}_{\nu}(s) {\cal{I}}_M(s) s ds} \ ,
\label{eq: defwstar}
\end{equation}

\begin{equation}
w_{DM}(c) = \frac{c^2}{f^2(c)} \int_{0}^{\infty}{\frac{\ln{(1 + c
y)} - c y (1 + cy)^{-1}}{(1 + cy)^2} dy} \ , \label{eq: defwdm}
\end{equation}

\begin{eqnarray}
w_{\star}^{DM}(c, n, R_e/r_s) & = & \frac{1}{{\cal{I}}_M(\infty)
f(c)} \nonumber \\
~ & \times & \int_{0}^{\infty}{{\cal{I}}_{\nu}(s) f(s, R_e/r_s) s
ds} \ , \label{eq: defwstardm}
\end{eqnarray}

\begin{eqnarray}
w_{DM}^{\star}(c, n, R_e/r_s) & = & \frac{1}{{\cal{I}}_M(\infty)
f(c)} \left ( \frac{R_e}{r_s} \right )^2 \nonumber
\\
~ & \times & \int_{0}^{\infty}{\left (1 + \frac{R_e}{r_s} s \right
) {\cal{I}}_{M}(s) s ds} \ . \label{eq: defwdmstar}
\end{eqnarray}
It is worth noting that, while $n$ and $R_e$ directly enter the
integrals above, $\Upsilon_{\star}$ and $I_e$ only work as scaling
parameters through $M_{\star}^T$. Moreover, a qualitative analysis
shows that the first term in Eq.(\ref{eq: defwpar}) turns out to
be the dominating one so that $w({\bf p})$ is essentially a
function of $n$ only.

\subsection{Scaling relations from the virial theorem}

Inserting Eqs.(\ref{eq: endk}) and (\ref{eq: endw}) into
Eq.(\ref{eq: virtheo}) and solving with respect to $\sigma_0$, we
get\,:

\begin{displaymath}
\sigma_0^2 = \frac{G M_{\star}^T}{R_e} \frac{w({\bf p})}{k({\bf
p})} \ .
\end{displaymath}
It is then convenient to introduce\,:

\begin{equation}
\langle I_e \rangle \equiv  \frac{L_T/2}{\pi R_e^2} \label{eq:
defieav}
\end{equation}
so that the total stellar mass is $M_{\star}^T = \Upsilon_{\star}
L_T = 2 \pi \Upsilon_{\star} \langle I_e \rangle R_e^2$ and the
relation above can be recast as\,:

\begin{equation}
2 \log{\sigma_0} = \log{\langle I_e \rangle} + \log{R_e} +
\log{\frac{w({\bf p})}{k({\bf p})}} + \log{(2 \pi G
\Upsilon_{\star})} \ . \label{eq: presvh}
\end{equation}
Let us suppose for a while that it is possible to neglect the dark
halo component. Should this be the case, both $k({\bf p})$ and
$w({\bf p})$ turn out to be a function of $(n, R_e, \langle I_e
\rangle)$, where hereafter we use $\langle I_e \rangle$ rather
than $I_e$ as a parameter\footnote{See, e.g. Graham \& Driver
(2005), for the relation between $I_e$ and $\langle I_e \rangle$
and other related formulae.}. Although determining how the ratio
$w({\bf p})/k({\bf p})$ depends on $(n, R_e, \langle I_e \rangle)$
needs for a full computation of the integrals involved, we can, as
first approximation, suppose that $w({\bf p})/k({\bf p})$ is
linear in a logarithmic scale. It is possible, therefore, to
write\,:

\begin{equation}
\log{\left [ w({\bf p})/k({\bf p}) \right ]} \simeq a \log{\langle
I_e \rangle} + b \log{R_e} + c \log{(n/4)} + d \ . \label{eq:
linearratio}
\end{equation}
From stellar population synthesis models, we know that the stellar
$M/L$ ratio may be correlated with the total stellar luminosity
$L_T$. Approximating this relation as a power\,-\,law, we can
therefore write\,:

\begin{equation}
\log{\Upsilon_{\star}} \simeq \alpha + \beta \log{L_T} = \alpha +
\beta \log{(2 \pi \langle I_e \rangle R_e^2)} \ . \label{eq:
linearups}
\end{equation}
Inserting Eqs.(\ref{eq: linearratio}) and (\ref{eq: linearups})
into Eq.(\ref{eq: presvh}), one finally gets\,:

\begin{equation}
\log{\sigma_0} = a_T \log{\langle I_e \rangle} + b_T \log{R_e} +
c_T \log{(n/4)} + d_T \label{eq: svh}
\end{equation}
with

\begin{equation}
\left \{
\begin{array}{l}
a_T = \displaystyle{\frac{a + \beta + 1}{2}} \\
~ \\
b_T = \displaystyle{\frac{b + 2 \beta + 1}{2}} \\
~ \\
c_T = \displaystyle{\frac{c}{2}} \\
~ \\
d_T = \displaystyle{\frac{\alpha}{2} + \frac{\beta + 1}{2} \log{(2
\pi)} + \frac{1}{2} \log{G} + d}
\end{array}
\right . \ . \label{eq: svhpar}
\end{equation}
As a first remark, let us note that the exact value of $d$ depends
on the adopted units. In the following, we will express $\sigma_0$
in km/s, $\langle I_e \rangle$ in ${\rm L_{\odot}/pc^2}$ and $R_e$
in kpc. In particular, having expressed $R_e$ in linear rather
than angular units makes $d$ dependent on the distance to the
galaxy. A second important caveat is related to our starting
hypothesis of having neglected the dark halo component. Actually,
we do know that galaxies are embedded in their dark matter haloes.
As a consequence, $w({\bf p})/k({\bf p})$ is a function of the
halo parameters too. To take into account this dependence, we
still assume Eq.(\ref{eq: linearratio}), but let $d$ be an unknown
function of $(c, M_{vir})$ to be determined by the data.

With all these caveats in mind, Eq.(\ref{eq: svh}) defines an
hyperplane in the logarithmic space allowing to express the
kinematic quantity $\log{\sigma_0}$ as a function of the
photometric parameters $\log{\langle I_e \rangle}$, $\log{R_e}$,
$\log(n/4)$. Since we have recovered it for a Sersic model under
the hypothesis of virial equilibrium, we will call it the {\it
Sersic Virial Hyperplane}.

Should our assumptions hold for real elliptical galaxies, the
Sersic Virial Hyperplane (hereafter SVH) should represent a tight
scaling relations among kinematic and photometric parameters. For
an idealized sample of galaxies perfectly satisfying our working
hypotheses and all at the same distance, the scatter around this
hyperplane should be generated by essentially two terms. First,
the halo parameters $(c, M_{vir})$ may differ on a
case\,-\,by\,-\,case basis. This is the same as saying that the
dark matter content in the inner regions or, equivalently, the
global $M/L$ ratio (defined as $M_{tot}/L_T$ rather than
$M_{\star}^T/L_T$)is different from one galaxy to another. On the
other hand, $d_T$ in Eq.(\ref{eq: svhpar}) may scatter from one
galaxy to another because of different values of the parameters
$(\alpha, \beta)$ of the $\Upsilon_{\star} - L_T$ relation because
of different details of the stellar evolution process. Note that
this latter effect may also affect the coefficients $(a_T, b_T)$
thus further increasing the scatter on the SVH.

It is worth stressing that the SVH may be seen as a generalization
of both the FP and PhP which, from this point of view, reduce to
particular cases of the SVH. Indeed, forcing the de Vaucouleurs
model to fit the galaxies surface brightness profiles is the same
as setting $n = 4$ in Eq.(\ref{eq: svh}). Solving with respect to
$\log{R_e}$, we get\,:

\begin{equation}
\log{R_e} = a_{FP} \log{\sigma_0} + b_{FP} \log{\langle I_e
\rangle} + c_{FP} \label{eq: fpsvh}
\end{equation}
which is indeed the FP with

\begin{equation}
\left \{
\begin{array}{l}
a_{FP} = 1/b_T \\
~ \\
b_{FP} = -a_T/b_T \\
~ \\
c_{FP} = -d_T/b_T \\
\end{array}
\right . \ . \label{eq: fppar}
\end{equation}
Actually, we do not expect that the coefficients of the observed
FP are equal to what is predicted by Eq.(\ref{eq: fppar}) since,
assuming $n = 4$ in the fit, can bias the estimate of $(R_e,
\langle I_e \rangle)$ in a way that depends on what the true value
of $n$ is. Moreover, the departure of $n$ from $n = 4$ introduces
a further scatter which is not included in the above expression
for $c_{FP}$. Although these effects have to be carefully
quantified, it is nevertheless worth stressing that the FP turns
out to be only a projection of the SVH on the plane $\log{(n/4)} =
0$ so that its coefficients may be (at least, in principle)
predicted on the basis of physical considerations.

On the other hand, when solving Eq.(\ref{eq: svh}) with respect to
$\log{R_e}$, we can also neglect the dependence on
$\log{\sigma_0}$ assuming this latter is, in a rough
approximation, the same for all galaxies. We thus get\,:

\begin{equation}
\log{R_e} = a_{PhP} \log{\langle I_e \rangle} + b_{PhP}
\log{(n/4)} + c_{PhP} \label{eq: phpsvh}
\end{equation}
with

\begin{equation}
\left \{
\begin{array}{l}
a_{PhP} = -a_T/b_T \\
~ \\
b_{PhP} = -c_T/b_T \\
~ \\
c_{PhP} = (1/b_T) \log{\sigma_0} -d_T/b_T \\
\end{array}
\right . \ . \label{eq: phppar}
\end{equation}
which is indeed the PhP. Note that, since $\log{\sigma_0}$ is
obviously not the same for all galaxies, the scatter in the PhP
may then be easily interpreted as a scatter in $\log{\sigma_0}$
and hence in the kinematic structure of the galaxies.

\section{Testing the SVH}

The derivation of the SVH in Eq.(\ref{eq: svh}) relies on two main
hypothesis which are analytically formalized in Eqs.(\ref{eq:
linearratio}) and (\ref{eq: linearups}). While the relation
$\Upsilon_{\star} \propto L_T^{\beta}$ may be tested resorting to
stellar population synthesis models, checking the validity of
Eq.(\ref{eq: linearratio}) needs for a detailed computation of
$w({\bf p})/k({\bf p})$ as function of the photometric parameters
$(n, \langle I_e \rangle, R_e)$, the stellar $M/L$ ratio
$\Upsilon_{\star}$ and the halo parameters $(c, M_{vir})$.
Performing such a computation over a fine grid in this six
dimensional space is prohibitively expensive. We can however rely
on a different and more reliable approach. Rather than evaluating
$w({\bf p})/k({\bf p})$, we simulate a sample of galaxies with
given values of the above parameters and use the set of equations
in Sect.\,3 to compute $\sigma_0$ and the ratio $w({\bf p})/k({\bf
p})$. We thus end up with a sample of {\it simulated measurements}
of $(\sigma_0, n, \langle I_e \rangle, R_e)$ which we fit with the
SVH thus determining the coefficients $(a_{sim}, b_{sim}, c_{sim},
d_{sim})$. Let us then assume that our simulated sample have the
same distribution for the parameters $(n, \langle I_e \rangle,
R_e, \Upsilon_{\star})$ as a real sample. Let us then denote with
$(a_{obs}, b_{obs}, c_{obs}, d_{obs})$ the values obtained by
fitting the SVH to the observed sample. Should our assumptions be
satisfied, we have to find\,:

\begin{equation}
\left \{
\begin{array}{l}
a_{obs} = a_{sim} + \beta/2 \\
~ \\
b_{obs} = b_{sim} + \beta \\
~ \\
c_{obs} = c_{sim} \\
~ \\
d_{obs} = d_{sim} + \alpha/2 + (\beta/2) \log{(2 \pi)} \\
\end{array}
\right . \nonumber \label{eq: obssim}
\end{equation}
with $\beta$ an estimate of the slope of the $\Upsilon_{\star} -
L_T$ relation. Moreover, the scatter in the observed SVH should be
the same as the one evaluated by the simulated sample. Actually,
since we do not have a model independent guess for $\beta$, we can
estimate $\beta$ as $2(a_{obs} - a_{sim})$ or as $b_{obs} -
b_{sim}$. The two estimates thus obtained should of course be
equal thus giving a further check on the validity of the model.

\subsection{The data}

In order to carry on the approach detailed above, a key step is a
sample of ETGs as large as possible. To this aim, we have started
from the NYU Value\,-\,Added Galaxy Catalog (hereafter, VAGC)
which is a cross\,-\,matched collection of galaxy catalogs
maintained for the study of galaxy formation and evolution
\cite{VAGC} and mainly based on the SDSS data release 4
\cite{DR4}. Among the vast amount of available data, we use the
{\it low\,-\,redshift} (hereafter, lowZ) catalog of galaxies with
estimated comoving distances in the range $10 < d < 150 \ h^{-1}
{\rm Mpc}$. We refer the reader to Blanton et al. (2005) and the
VACG website\footnote{{\tt http://sdss.physics.nyu.edu/vagc/}} for
details on the compilation of the catalog\footnote{Note that the
version we are using is updated only to the second SDSS data
release \cite{DR2} covering an effective survey area of 2220.9
square degrees.}.

We shortcut the lowZ catalog only retaining those data we are
mainly interested in and rejecting all the galaxies with no
measurements of $\sigma_0$ leaving us with 24387 out of 28089
objects with magnitudes in the five SDSS filters $u' g' r' i' z'$.
In order to select only ETGs, we apply a set of criteria which we
briefly details below.

\begin{enumerate}
\item{A Sersic profile has been fitted to the surface brightness profile of each galaxy
using an automated pipeline retrieving the parameters $(n, R_e,
A)$ with $R_e$ in $arcsec$ and $A$ the total flux in nanomaggies.
As a first criterium, we select only galaxies with $2.5 \le n \le
5.5$, where the upper limit is dictated by the code limit $n =
6.0$. This selection is performed using the fit in the $i'$ band
since it is less affected by dust without the reduced efficiency
of the $z'$ band.}

\item{As a second criterium, we impose the cut $R_{90}/R_{50} > 2.6$ \cite{Shima01}
with $R_{90}$ and $R_{50}$ the Petrosian radii containing $90\%$
and $50\%$ of the total luminosity as estimated by the data. Note
that, although the Petrosian radii do non depend on any fitting,
the ratio $R_{90}/R_{50}$ is correlated with $n$ so that the two
criteria are somewhat redundant. Removing one of them or changing
the order does not alter in a significant way the final sample.}

\item{We exclude all galaxies with $\sigma_0 < 70 \ {\rm
km/s}$ since the dispersion velocity measurements for these
systems may be problematic \cite{B05}.}

\item{Elliptical galaxies are segregated in a well defined region of the color\,-\,magnitude
plane. In order to further restrict our sample, we therefore
impose the cut $(g - r)_{-} \le g - r \le (g - r)_{+}$ with $(g -
r)_{\pm} = p M_r + q \pm \delta$. Here, $M_r$ is the absolute
magnitude in the $r$ filter and the parameters $(p, q, \delta)$
have been tailored from Fig.\,2 in Bernardi et al. (2005) where a
different ETG sample has been extracted from the SDSS DR2.}

\end{enumerate}
The final sample thus obtained contains 5142 galaxies out of an
initial catalog containing 24387 objects. It is worth noting that
most of the rejected objects have been excluded by the first three
cuts (retaining only 5172 entries), while the fourth cut only
removes 30 further galaxies. This is reassuring since the last cut
is somewhat qualitative and based on a different set of selection
criteria\footnote{It is worth noting that we cannot use these
criteria since the lowZ catalog does not report the parameters
which the selection by Bernardi et al. (2005) are based on.}
\cite{B05}.

We then use the data reported in the lowZ catalog for the galaxies
in the above sample to collect the quantities listed below.

\begin{itemize}
\item{{\it Photometric quantities.} While the Sersic index $n$ and
the effective radius $R_e$ in $arcsec$ are directly available in
the lowZ catalog, the average effective intensity $\langle I_e
\rangle$ is not present. To this aim, we first convert the total
flux $A$ (in nanomaggies) reported in the catalog in the apparent
total magnitude $m_t$ as \cite{VAGC}\,:

\begin{displaymath}
m_t = 22.5 - 2.5 \log{A} \ .
\end{displaymath}
We then use the assumed concordance cosmological model to estimate
the total absolute magnitude ${\cal{M}}_t$ as\,:

\begin{eqnarray}
{\cal{M}}_t & = & m_t - 5 \log{D_L(z)} + 5 \log{h} - 10 \log{(1 +
z)} \nonumber \\
~ & - & K(z) - A_{G} - 42.38 \nonumber \label{eq: mtest}
\end{eqnarray}
where $z$ is the galaxy redshift, $D_L$ the luminosity distance,
$K(z)$ the $K$\,-\,correction, $A_G$ the galactic extinction, and
the term $10 \log{(1 + z)}$ takes into account the cosmological
dimming. While $K(z)$ and $A_G$ are reported in the catalog for
each of the five SDSS filters, our use of the luminosity distance
makes the estimate of $M_t$ cosmological model dependent. However,
for the values of $z$ involved, the dependence on the cosmological
model is actually meaningless. We finally estimate\,:

\begin{equation}
\langle I_e \rangle = 10^{-6} \times \frac{{\rm dex}[({\cal{M}}_t
- {\cal{M}}_{\odot})/2.5]}{2 \pi R_e^2} \label{eq: ieest}
\end{equation}
with ${\rm dex}(x) \equiv 10^{x}$, ${\cal{M}}_{\odot}$ the Sun
absolute magnitude in the given filter\footnote{We use
${\cal{M}}_{\odot} = (5.82, 5.44, 4.52, 4.11, 3.89)$ for the $u'
g' r' i' z'$ filters respectively as evaluated from a detailed Sun
model reported in {\tt www.ucolick.org/$\sim$cnaw/sun.html}}. We
stress that, in Eq.(\ref{eq: ieest}), $R_e$ is expressed in $kpc$
rather than $arcsec$. To this aim, we simply use\,:

\begin{displaymath}
R_e({\rm kpc}) = R_e(arcsec) \times D_A(z)/206265
\end{displaymath}
with $D_A(z)$ the angular diameter distance in Mpc.}

\item{{\it Kinematic quantities.} The lowZ catalog reports the
velocity dispersion and its error as determined from the SDSS
spectrum of the galaxy. This is measured in a circular aperture of
fixed radius $R_{ap} = R_{SDSS} = 1.5 \ arcsec$, while $\sigma_0$
in Eq.(\ref{eq: sigmazero}) has been estimated for $R_{ap} =
R_e/8$. To correct for this offset, we follow J$\o$rgensen et al.
(1995, 1996) setting\,:

\begin{equation}
\sigma_0^{obs} = \sigma_0^{lowZ} \times \left (
\frac{R_{SDSS}}{R_e/8} \right )^{0.04} \label{eq: sigmacorr}
\end{equation}
with $\sigma_0^{lowZ}$ the value in the catalog and $R_e$ in
$arcsec$ here.}

\item{{\it Auxiliary quantities.} The lowZ catalog contains a wealth
of information on each object that we really do not need for our
analysis. We do, however, add to our catalog some further
quantities that we will use for check. In particular, we include
the absolute magnitude ${\cal{M}}_{SDSS}$ as estimated from the
image rather than the fit, and the average effective surface
brightness computed as \cite{GD05}\,:

\begin{displaymath}
\langle \mu_e \rangle = \langle \mu_e \rangle_{abs} + 10 \log{(1 +
z)} + K(z) + A_G
\end{displaymath}
with

\begin{displaymath}
\langle \mu_e \rangle_{abs} = {\cal{M}}_t + 2.5 \log{(2 \pi
R_e^2)} + 36.57
\end{displaymath}
with $R_e$ in kpc. Note that $\langle \mu_e \rangle$ rather than
$\log{\langle I_e \rangle}$ is often used in the FP and PhP fit.}

\end{itemize}
Although the Sersic law is known to well fit the surface
brightness profile of ETGs, it is worth noting that our derivation
of $\langle I_e \rangle$ relies on extrapolating the fit results
well beyond the visible edge of the galaxy. As such, it is
possible that ${\cal{M}}_t$ provides a biased estimate of the
actual total absolute magnitude of the galaxy which is better
represented by ${\cal{M}}_{SDSS}$. A bias in the estimate of
${\cal{M}}_t$ propagates on the estimates of the colors which may
be related to the stellar mass by population synthesis models. In
order to reduce as more as possible such a bias, we have studied
the histogram of $\Delta col = col_{obs} - col_{est}$, where
$col_{obs} = {\cal{M}}_{SDSS, j} - {\cal{M}}_{SDSS, k}$ and
$col_{est} = {\cal{M}}_{t, j} - {\cal{M}}_{t, k}$. In principle,
all these histograms should be centered at the null value with a
small scatter. After removing 42 outliers, this is indeed the case
for $g' - r'$ and $r' - i'$ (with rms values of 0.08 and 0.06 mag
respectively), while this is not for $u' - r'$ and $i' - z'$
(having rms values of 0.21 and 0.13 mag). Motivated by this bias,
we will use only data in $g'r'i'$ filters when fitting the SVH
taking, in particular, the results from the fit to the $i'$ band
as the fiducial ones.

\subsection{The simulated sample}

The approach we have outlined above to test the validity of the
SVH (and, hence, of the hypotheses it relies on) is the
construction of simulated sample of ETG which is as similar as
possible to the real one. In particular, the photometric
parameters of the simulated sample should match as closely as
possible those of the observed one.

To this aim, we first look at the histograms of the Sersic
parameters $(n, R_e", \log{\langle I_e \rangle})$ with $R_e"$ the
effective radius in $arcsec$, the effective radius $R_e$ in kpc,
and the distance modulus $dm$ computed as\,:

\begin{eqnarray}
dm & = & 5 \log{D_L(z)} - 5 \log{h} + 10 \log{(1 + z)} \nonumber
\\ ~ & + & K(z) + A_{G} + 42.38 \ . \nonumber
\end{eqnarray}
Note that all the quantities entering the definition above are
available for each galaxy in the lowZ catalog so that the
distribution of $dm$ values is easily obtained. We use the $i'$
values as fiducial ones and start from the histogram thus obtained
to compute the cumulative distribution functions (CDF) for the
quantities of interest. A simulated galaxy is then generated
according to the steps we sketch below.

\begin{enumerate}
\item[i.]{Using the CDFs obtained above, we generate the $i'$ values of the parameters
$(n, R_e", \log{\langle I_e \rangle}, R_e, dm)$. It is worth
stressing that the ratio $R_e"/R_e$ provide an estimate of the
angular diameter distance $D_A(z)$ to the simulated galaxy and,
hence, of the redshift provided a cosmological model has been set.
As an alternative approach, one could generate the redshift $z$
and $R_e$ to infer $R_e"$. Although the two approaches are in
principle equivalent, we have checked that our choice is more
stable from a computational point of view. Note also that
generating $dm$ directly makes it possible to avoid deriving an
estimate for $K(z)$ and $A_G$ thus reducing the number of
quantities to be generated.}

\item[ii.]{The parameters we have generated are given in the $i'$ filter\footnote{
Note that $dm$ depends on the filter used because of the terms
$K(z)$ and $A_G$ that are wavelength dependent.}, while we would
like to simulate the galaxy properties in all the observed
filters. To this aim, we have verified that linear relations
between the parameters in different filters exist in the real
sample so that we can write \,:

\begin{displaymath}
{\bf p}_f = {\cal{A}} {\bf p}_{i'}
\end{displaymath}
with ${\bf p}_f$ the set of parameters $(n, R_e", \log{\langle I_e
\rangle}, R_e, dm)$ in the $f$ filter and ${\cal{A}}$ a diagonal
matrix. Using this relations, we then generate the values of ${\bf
p}_f$ for the simulated galaxy in the remaining $u' g' r' z'$
filters. We are now able to compute all the described photometric
and auxiliary quantities in the case of the real sample.}

\item[iii.]{As a last step, we need to compute the velocity dispersion $\sigma_0$
and its error considering the ETG modelling described in Sect.\,2.
To this aim, we need to generate three further quantities, namely
the stellar $M/L$ ratio $\Upsilon_{\star}$ and the NFW parameters
$(c, M_{vir})$. As a first step, we construct the CDF for
$\Upsilon_{\star}$ from the real sample. To this aim, for each
observed galaxy, we estimate \cite{FHP98}\,:

\begin{equation}
\Upsilon_{\star}^V = 4.0 + 0.38 \left [ t(z) - 10 \right ]
\label{eq: mtoltg}
\end{equation}
with $t(z)$ the age (in Gyr) of the galaxy at redshift $z$
computed assuming a formation redshift $z_F = 2$. We then convert
to the $i'$ band $M/L$ as\,:

\begin{equation}
\log{\Upsilon_{\star}} = \log{\Upsilon_{\star}^V} + 0.4 \left [
\left ( V - i' \right ) - \left ( V - i' \right )_{\odot} \right ]
\end{equation}
with $V - i' = 0.79$ \cite{FSI95}. The CDF thus obtained is then
used as seed for the random generation of the simulated galaxy
$\Upsilon_{\star}$. We can then estimate the total stellar mass
$M_{\star}^T$ as $2 \pi \Upsilon_{\star} R_e^2 \langle I_e
\rangle$. In order to set the halo model parameters, we first
generate the fraction of the total galaxy mass represented by the
dark matter. Defining $\eta = M_{vir}/(M_{\star}^T + M_{vir})$, we
extract $\eta$ from a Gaussian distribution centred on $0.90$ and
with dispersion 0.03. It is then a simple algebra to find out
$M_{vir} = \eta/(1 - \eta) \times M_{\star}^T$ so that we have
only to choose a value for the concentration $c$. To this aim, we
use Eq.(\ref{eq: cmvir}) to set the centre of a Gaussian
distribution for $c$ with a dispersion of 0.11 dex. Since the
model is then fully assigned, we may resort to Eq.(\ref{eq:
sigmazero}) to estimate $\sigma_0$ and to Eqs.(\ref{eq: endk}) and
(\ref{eq: endw}) to compute $k({\bf p})$ and $w({\bf p})$. The
error on the predicted $\sigma_0$ is finally set as $\varepsilon
\sigma_0$ with $\varepsilon$ extracted from a CDF obtained by the
same distribution as the observed one.}

\end{enumerate}
We repeat the procedure outlined above ${\cal{N}} \simeq 10000$
times thus obtaining a simulated sample to which we apply the same
selection procedure as for the real data. We finally end up with a
simulated sample containing (approximately) the same number of
ETGs as the real one. While, by construction, the distribution of
the photometric parameters $(n, R_e, \langle I_e \rangle)$ is the
same as the observed one in the $i'$ filter, we have checked that
the same applies in the other filters. Moreover, we have checked
that the statistical distributions of $m_t$ and ${\cal{M}}_t$ and
of the colors are the same as the observed ones. We are therefore
confident that the simulated sample is a fair representation of
the real data thus making sense to use it as input for testing the
SVH as described above.

\section{Results}

The real and the simulated samples described above are the input
ingredients for the procedure used to test the SVH. As a
preliminary step, however, we have to choose a method to fit the
SVH to a given dataset. As well known \cite{FB92,AB96,LaB00},
fitting a linear relation in a multiparameter space may seriously
depend on the method adopted to get the estimate of the
coefficients. Moreover, the choice of the most suitable method
also depends on the uncertainties on the model parameters. In the
present case, there is a further complication since the lowZ
catalog does report the measurement uncertainties on $\sigma_0$,
but not on the photometric parameters. As such, we have therefore
decided to consider $\sigma_0$ as the dependent variable and
neglect, in a first approximation, the measurement errors. We
therefore use a {\it direct fit} approach, i.e. we minimize\,:

\begin{equation}
\chi^2 = \sum_{i = 1}^{{\cal{N}}_{obs}}{\left [ \left (
\log{\sigma_0} \right )_{i, obs} - \left ( \log{\sigma_0} \right
)_{i, SVH} \right ]^2} \label{eq: chi2}
\end{equation}
where $\left ( \log{\sigma_0} \right )_{i, obs}$ and $\left (
\log{\sigma_0} \right )_{i, est}$ are respectively the observed
value and the predicted one through the SVH (\ref{eq: svh}) for
the i\,-\,th galaxy and the sum is over the ${\cal{N}}_{obs}$
elements in the (observed or simulated) sample. We estimate the
parameters $(a, b, c, d)$ by minimizing the $\chi^2$ defined above
with an iterative 1.5$\sigma$\,-\,clipping rejection
scheme\footnote{The $\sigma$\,-\,clipping procedure retains more
than $85\%$ of the data. The direct fit with no
$\sigma$\,-\,clipping leads to consistent estimates of the SVH
parameters, but the scatter turns out to be artificially increased
because of clear outliers.}. The root mean square of the fit
residuals is then taken as an estimate of the intrinsic scatter
around the SVH.

An important remark is in order. Although one may question on what
fitting procedure is the most suitable one, we stress that we are
here mainly interested in testing the consistency between what
data effectively mean and what our theoretical assumptions
predict. To this aim, it is more important that both the observed
and the simulated samples are examined in the same way rather than
discussing on which method is the best one. We are therefore
confident that the choice of the direct fit approach does not
alter the main conclusions we will draw from the comparison among
the observed and predicted SVH coefficients and scatter.

As a final warning, note that in the rest of this section, we only
consider data and simulations carried out in the $i'$ filter in
order to avoid any bias from our qualitative generation of the
galaxy properties in other filters. We have however checked that
the main results are unchanged should we use the $g'$ or $r'$
filters. Moreover, we only report numerical values of one fiducial
simulated sample. However, averaging over different simulations
has no significant effect on the results.

\subsection{The linear relation hypothesis}

As a first fundamental hypothesis in deriving the SVH from the
virial theorem, we have assumed that the ratio $w({\bf p})/k({\bf
p})$ is log\,-\,linear in the photometric parameters $(n, R_e,
\langle I_e \rangle)$. With the simulated sample at hand, we may
check whether this is indeed the case. To this aim, we have used
the direct fit approach fitting Eq.(\ref{eq: linearratio}) to the
values of $\log{[w({\bf p})/k({\bf p}]}$ computed for the galaxies
in the simulated sample. We find a linear relation among $y =
\log{[w({\bf p})/k({\bf p})]}$ and $x = a \log{\langle I_e
\rangle} + b \log{R_e} + c \log{(n/4)} + d$ thus validating
Eq.(\ref{eq: linearratio}). For the simulation reported, we
find\footnote{We are not concerned now with the value of $d$ since
this also depends on the average distance of the galaxies
sample.}\,:

\begin{displaymath}
(a, b, c) \simeq (-0.05, 0.04, 0.29)
\end{displaymath}
with an intrinsic scatter $\sigma_{int} \simeq 0.08$. Using such
values and Eq.(\ref{eq: svhpar}) with $\beta = 0$, we then
estimate\,:

\begin{displaymath}
(a_T, b_T, c_T) \simeq (0.48, 0.52, 0.14) \ .
\end{displaymath}
Should our ETG modelling and the virial theorem apply, we then
expect that the SVH in Eq.(\ref{eq: svh}) with $(a, b, c)$ fits
the observed data reasonably well. On the one hand, we have not
used any $\Upsilon_{\star} - L_T$ relation (that is why we have
set $\beta = 0$ above) when computing $w({\bf p})/k({\bf p})$ and
$\sigma_0$. However, stellar population synthesis (SPS) models
suggest that such a relation indeed exists. One can, in principle,
use a SPS code to get an estimate of $(\alpha, \beta)$, but this
should introduce a further set of assumptions (such as, e.g., star
formation history, metallicity, internal absorption) needed as
input for the code itself. In order to escape these problems, we
have therefore decided to follow a different strategy solving
Eqs.(\ref{eq: obssim}) with respect to $(\alpha, \beta)$, using as
input the observed and predicted values of the SVH parameters.
Note that, since $\beta$ may be solved from two different
equations, this will offer also a consistency check for the
theory.

\subsection{The SVH coefficients}

\begin{figure}
\includegraphics[width=8.5cm]{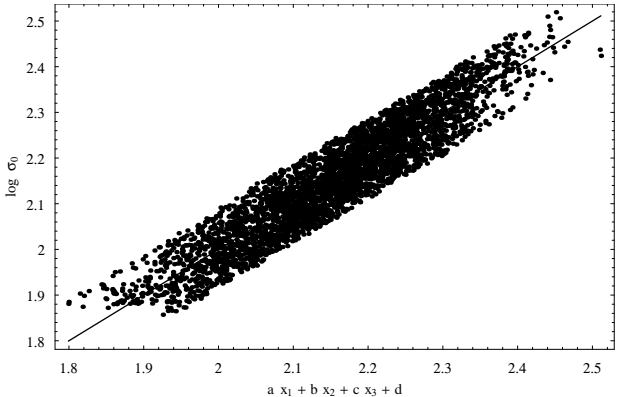}
\caption{On the $y$\,-\,axis, $\log{\sigma_0}$ is reported as
observed for the ETG sample. On the $x$\,-\,axis, the same
quantity is reported as fitted by the linear relation $a x_1 + b
x_2 + c x_3 + d$ with $x_1 = \log{\langle I_e \rangle}$, $x_2 =
\log{R_e}$, $x_3 = \log{(n/4})$. The coefficients $(a, b, c, d)$
are estimated by a direct fit as explained in the text.}
\label{fig: svhobs}
\end{figure}

In order to test definitively our hypotheses, we first fit the SVH
to the observed data using the direct fit approach outlined above.
The result is shown in Fig.\,\ref{fig: svhobs} where we plot $y$
vs $x$ with $y = \log{\sigma_0}$ and $x$ as above, finding

\begin{equation}
(a_{obs}, b_{obs}, c_{obs}, d_{obs}) = (0.493, 0.640, 0.145,
0.754) \label{eq: svhparobs}
\end{equation}
with an intrinsic scatter $\sigma_{int}^{obs} = 0.045$. As
Fig.\,\ref{fig: svhobs} clearly shows, the SVH fits the data
remarkably well over the full observed range in $\log{\sigma_0}$.
The fit residuals are quite small and do not correlate neither
with $x$ nor with $y$. We can therefore safely conclude that the
SVH is indeed observationally well founded\footnote{It is worth
noting that a first attempt to fit a hyperplane to a small sample
of Coma and Fornax galaxies (collected from the literature
available at that time) has yet been successfully performed in
Graham (2005). However, in that work, the fit was only empirically
motivated and the rest of the paper gives off the dependence on
$\sigma_0$ concentrating on the PhP. We can therefore consider the
present work as the first theoretically motivated study of the SVH
based on a large and homogenous ETG sample.}.

\begin{figure}
\includegraphics[width=8.5cm]{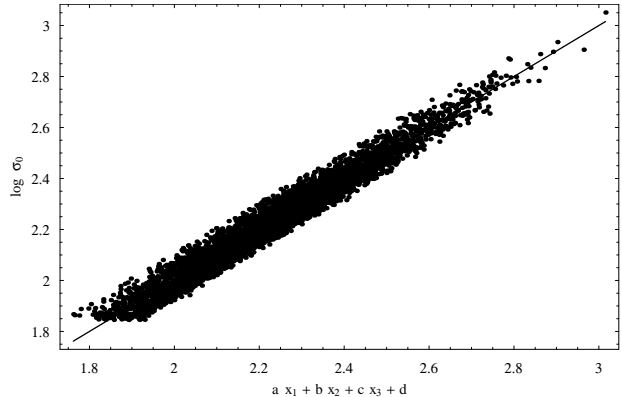}
\caption{On the $y$\,-\,axis, we report $\log{\sigma_0}$ as
computed for the simulated ETG sample. On the $x$\,-\,axis, we
instead report the same quantity as fitted by the linear relation
$a x_1 + b x_2 + c x_3 + d$ with $x_1 = \log{\langle I_e
\rangle}$, $x_2 = \log{R_e}$, $x_3 = \log{(n/4})$. The
coefficients $(a, b, c, d)$ are estimated from a direct fit as
explained in the text.} \label{fig: svhsim}
\end{figure}

Motivated by these encouraging result, we now turn to the
simulated sample repeating the same fit as above, but with $y$ now
computed on the basis of our ETG modelling and assumptions on the
halo parameters. Fig.\,\ref{fig: svhsim} clearly shows that the
SVH fits very well the simulated data with\,:

\begin{equation}
(a_{sim}, b_{sim}, c_{sim}, d_{sim}) = (0.475, 0.515, 0.142,
0.822) \label{eq: svhparsim}
\end{equation}
with an intrinsic scatter $\sigma_{int}^{sim} = 0.040$. Comparing
Eq.(\ref{eq: svhparobs}) and (\ref{eq: svhparsim}) is a rewarding
task. Indeed, $c_{obs}$ and $c_{sim}$ are remarkably close which
is exactly what we do expect because of Eq.(\ref{eq: svhpar}).
Moreover, the scatter is quite similar with a modest underestimate
($\Delta \sigma_{int} = \sigma_{int}^{obs} - \sigma_{int}^{sim} =
0.05$). Considering that we have included as source of the scatter
only the one due to the $c - M_{vir}$ relation and neglected that
introduced by the conversion from $\Upsilon_{\star}^V$ to
$\Upsilon_{\star}$, such a small value of $\Delta \sigma_{int}$
has not to be considered as a shortcoming of our theoretical
assumptions.

Solving Eq.(\ref{eq: obssim}) with respect to $\beta$ using the
values of $a_{obs}$ and $a_{sim}$ gives $\beta \simeq 0.03$, while
$\beta \simeq 0.12$ is obtained from $b_{obs}$ and $b_{sim}$.
These two estimates are not consistent with respect to each other
so that one could be tempted to conclude that something is wrong
with our hypotheses. There are however some important systematic
errors which these difference could be ascribed to. First, we
stress that the values in Eq.(\ref{eq: svhparsim}) have been
obtained from a simulated ETG sample obtained under the
simplifying hypothesis of spherical symmetry. As well known
\cite{BT87}, for a given total mass, the velocity dispersion may
considerably differ between spherically and flattened systems. As
a second simplification, we have computed the kinetic energy in
the case of isotropy in the velocity space, while it is likely
that ETGs are anisotropic systems. Quantifying these effects is
not possible unless one takes carefully into account the
distribution in the intrinsic flattening $q$ and the anisotropy
parameter $\beta_{\sigma}$ which are largely unknown. We can
roughly quantify the effect of these systematics errors by first
writing the {\it true} velocity dispersion as\,:

\begin{displaymath}
\sigma_{0, true}^2 = \sigma_0^2 + \sigma_{0, sys}^2 = \sigma_0^{2}
\times \left [ 1 + \left ( \frac{\sigma_{0, sys}}{\sigma_0} \right
)^2 \right ]
\end{displaymath}
with $\sigma_{0, sys}$ the term we are neglecting due to the
systematic errors above. Using the SVH for $\sigma_0^2$, we get\,:

\begin{eqnarray}
\log{\sigma_{0, true}} & = & a_T \log{\langle I_e \rangle} + b_T
\log{R_e} + c_T \log{(n/4)} + d_T \nonumber \\
~ & + &  \frac{1}{2} \log{\left [ 1 + \left ( \frac{\sigma_{0,
sys}}{\sigma_0} \right )^2 \right ]} \ . \label{eq: svhcorr}
\end{eqnarray}
The fit to the real data is consistent with the existence of the
SVH so that we can argue that the last term on the r.h.s. is
linear in logarithmic units. In particular, by writing

\begin{equation}
\log{\left [ 1 + \left ( \frac{\sigma_{0, sys}}{\sigma_0} \right
)^2 \right ]} = \delta \log{\langle I_e \rangle} \label{eq:
svhcorrterm}
\end{equation}
the SVH is recovered provided that we replace $a_T$ with $a_T +
\delta/2$. The matching between simulated and observed
coefficients is now obtained for\,:

\begin{equation}
\left \{
\begin{array}{l}
a_{sim} + \beta/2 + \delta = a_{obs} \\
~ \\
b_{sim} + \beta = b_{obs} \\
\end{array}
\right . \ . \nonumber
\end{equation}
From the values of $(a_{obs}, b_{obs}, a_{sim}, b_{sim})$, we thus
get\,:

\begin{table*}
\caption{Best fit coefficients and intrinsic scatter of the SVH in
the $g' r' i'$ filters obtained as described in the text.}
\begin{center}
\begin{tabular}{|c|c|c|c|c|c|}
\hline Filter & $a$ & $b$ & $c$ & $d$ & $\sigma_{int}$ \\
\hline $g'$ & $0.433 \pm 0.008$ & $0.595 \pm 0.010$ & $0.102 \pm
0.014$ & $0.880 \pm 0.022$ & $0.089 \pm 0.002$ \\
$r'$ & $0.460 \pm 0.008$ & $0.600 \pm 0.010$ & $0.113 \pm 0.014$ &
$0.841 \pm 0.022$ & $0.081 \pm 0.002$ \\ $i'$ & $0.469 \pm 0.008$
& $0.594 \pm 0.010$ & $0.146 \pm 0.014$ & $0.824 \pm 0.022$ &
$0.077 \pm 0.003$ \\\hline
\end{tabular}
\end{center}
\end{table*}

\begin{displaymath}
(\beta, \delta) \simeq (0.12, -0.09) \ .
\end{displaymath}
It is worth noting that such a small value of $\beta$ indicates a
quite weak correlation between $\Upsilon_{\star}$ and $L_T$. This
is qualitatively consistent with the finding of Padmanabhan et al.
(2004). Estimating the stellar $M/L$ from the correlation with the
$D_{4000}$ strength \cite{K03}, these authors find
$\Upsilon_{\star}$ to be almost constant with $L_T$ which compares
well with our estimated $\beta$. Moreover, defining the dynamical
mass $M_{dyn}$ as the total mass within (approximately) $R_e$ and
estimating it as $M_{dyn} = (1.65)^2 \sigma_0^2 R_e/G$,
Padmanabhan et al. finds $M_{dyn}/L_T \propto L_T^{0.17}$. With
our finding $\Upsilon_{\star} \propto L_T^{0.12}$, we can recover
the Padmanabhan et al. result provided $M_{dyn}/M_{\star}^T
\propto L_T^{-0.04}$. Using the real data and estimating
$\Upsilon_{\star}$ from Eq.(\ref{eq: mtoltg}), we find
$M_{dyn}/M_{\star}^T \propto L_T^{0.03}$ so that our model
predicts $M_{dyn}/L_T \propto L_T^{0.15}$ in remarkable agreement
with the Padmanabhan et al. value.

Notwithstanding the encouraging result for $\beta$, it is worth
stressing that Eq.(\ref{eq: svhcorrterm}) is actually an
unmotivated assumption. Nevertheless, we can qualitatively assess
whether such a hypothesis is reasonable by a qualitative analogy
with spiral galaxies. Some recent works \cite{R99,N07} claims that
the shape of the rotation curve is determined not only by the
total luminosity, but also weakly depends on the mean surface
density. Considering $\sigma_0$ as the analog of $v_c$ for ETGs,
we may qualitatively assume that the power\,-\,law relation
$\Upsilon_{\star} - L_T$ describes the dependence of $\sigma_{0,
true}$ on $L_T$, while Eq.(\ref{eq: svhcorrterm}) takes into
account a possible effect related to $\langle I_e \rangle$.
Although this interpretation is, strictly speaking, unmotivated,
the low $\delta$ value and the positive match with the Padmanabhan
et al. finding make us confident that Eq.(\ref{eq: svhcorrterm})
is not unrealistic.

Considering these encouraging results, we therefore conclude that
the theoretically predicted SVH can reasonably coincide with the
observational one.

\section{The observed SVH}

In the previous section, we were mainly interested in testing the
theoretical bases of SVH so that what is pressing is the need for
a consistent and homogenous analysis of both the observed and the
simulated data. Such a consideration has motivated our choice of
the simplified fitting method which is, however, not best suited
to derive the {\it true} SVH coefficients. In particular, we have
not taken into account the measurement uncertainties on $\sigma_0$
giving the same weight to each ETG in the sample. Moreover, the
scatter in the SVH has been estimated {\it a posteriori}, while
one should more correctly take into account such an intrinsic
scatter {\it a priori}, i.e. as a further parameter to be
determined by the fitting procedure.

The Bayesian probabilistic approach offers an ideal route to solve
this problem. We do not enter in any detail here referring the
interested reader to the vast literature available (see, e.g., D'
Agostini (2004) and refs. therein). Let us suppose that a linear
relation holds as\,:

\begin{equation}
y = a x_1 + b x_2 + c x_3 + d \label{eq: linear}
\end{equation}
and let $\sigma_{int}$ be its intrinsic scatter. If the errors on
the variables involved are statistically independent, one can
demonstrate that the best fit estimate of the parameters $(a, b,
c, d)$ and of the scatter $\sigma_{int}$ is obtained by minimizing
the following merit function \cite{Dago05}\,:

\begin{eqnarray}
- \ln{{\cal{L}}} & = & \displaystyle{\frac{1}{2} \sum_{i =
1}^{{\cal{N}}}{\ln{\left (\sigma_{int}^2 + \sigma_{y,i}^2 + a^2
\sigma_{1,i}^2 + b^2 \sigma_{2,i}^2 + c^2 \sigma_{3,i}^2 \right)}}} \nonumber \\
~ & + & \displaystyle{\frac{1}{2} \sum_{i =
1}^{{\cal{N}}}{\frac{(y_i - a x_{1,i} - b x_{2,i} - c x_{3,i} -
d)^2}{\sigma_{int}^2 + \sigma_{y,i}^2 + a^2 \sigma_{1,i}^2 + b^2
\sigma_{2,i}^2 + c^2 \sigma_{3,i}^2}}}  \label{eq: lnl}
\end{eqnarray}
with $\sigma_{y,i}$ and $\sigma_{j,i}$ the errors on $y$ and $x_j$
for the i\,-\,th object and the sum is over the ${\cal{N}}$
objects in the sample. It is worth stressing that the minimization
with respect to $d$ may be performed analytically, i.e., for given
$(a, b, c, \sigma_{int})$, the best fit $d$ is given by\,:

\begin{figure}
\includegraphics[width=8.5cm]{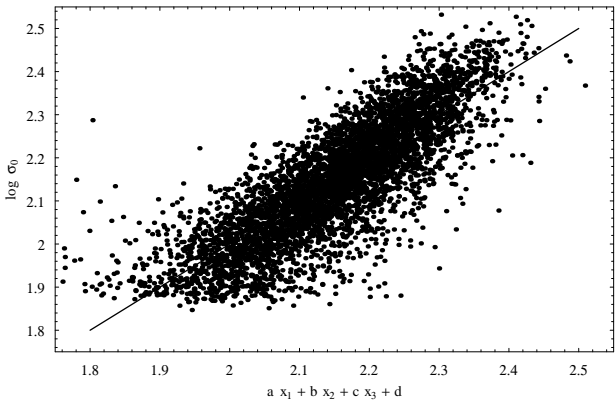}
\caption{Same as Fig.\,\ref{fig: svhobs} but with the fit
coefficients estimated from the Bayesian method.} \label{fig:
svhbayes}
\end{figure}

\begin{equation}
d = \frac{\sum_{i = 1}^{{\cal{N}}}{\frac{y_i - a x_{1,i} - b
x_{2,i} - c x_{3,i}}{\sigma_{int}^2 + \sigma_{y,i}^2 + a^2
\sigma_{1,i}^2 + b^2 \sigma_{2,i}^2 + c^2
\sigma_{3,i}^2}}}{\sum_{i = 1}^{{\cal{N}}}{\frac{1}{\sigma_{int}^2
+ \sigma_{y,i}^2 + a^2 \sigma_{1,i}^2 + b^2 \sigma_{2,i}^2 + c^2
\sigma_{3,i}^2}}} \ . \label{eq: dbf}
\end{equation}
With this value of $d$, one can compute $-\ln{{\cal{L}}}$ and then
find the set of parameters $(a, b, c, d, \sigma_{int})$ that
minimizes it, therefore representing the best fit solution.

\begin{table*}
\caption{Best fit coefficients and intrinsic scatter of the SVH as
rewritten in Eq.(\ref{eq: svhlogre}) in the $g' r' i'$ filters.}
\begin{center}
\begin{tabular}{|c|c|c|c|c|c|}
\hline Filter & $a_e$ & $b_e$ & $c_e$ & $d_e$ & $\sigma_{int,e}$ \\
\hline $g'$ & $0.897 \pm 0.020$ & $-0.659 \pm 0.007$ & $0.064 \pm
0.022$ & $0.052 \pm 0.039$ & $0.110 \pm 0.002$ \\
$r'$ & $0.973 \pm 0.022$ & $-0.678 \pm 0.008$ & $0.076 \pm 0.023$
& $-0.111 \pm 0.039$ & $0.100 \pm 0.002$ \\ $i'$ & $1.020 \pm
0.020$ & $-0.695 \pm 0.007$ & $0.052 \pm 0.021$ & $-0.189 \pm
0.036$ & $0.100 \pm 0.003$ \\\hline
\end{tabular}
\end{center}
\end{table*}

The general formulae (\ref{eq: linear})\,--\,(\ref{eq: dbf}) may
be easily adapted to our problem. Eqs.(\ref{eq: svh}) and
(\ref{eq: linear}) may be identified setting\,:

\begin{displaymath}
y = \log{\sigma_0} \ , \ x_1 = \log{\langle I_e \rangle} \ , x_2 =
\log{R_e} \ , x_3 = \log{(n/4)} \ .
\end{displaymath}
Note that, for the data at hand, $\sigma_1 = \sigma_2 = \sigma_3 =
0$. Finally, we find the results summarized in Table 1 for the SVH
in the $g' r' i'$ filters (excluding therefore the $u'$ and $z'$
data for the problems hinted to in Sect.\,4) where the errors on
the fit parameters have been estimated by 1000 bootstrap
resampling. Fig.\,\ref{fig: svhbayes} gives a visual impression of
the quality of the fit superimposing the best fit curve to the
$i'$ filter data.

Comparing the results in Table 1 for the $i'$ filter and those in
Eq.(\ref{eq: svhparobs}), we find a good agreement for both $c$
and $d$ coefficients, while a less good but still reasonable
matching is obtained for $a$ and $b$. From this point of view,
both the direct fit and the Bayesian approach seem to provide
reliable estimate of the SVH parameters. However, the scatter in
Table 1 is almost two times larger than the one estimated by the
direct fit method. This is an expected result. The two fitting
procedures actually minimizes a similar merit function and indeed,
in Sect. 5, we have looked for the minimum of Eq.(\ref{eq: lnl})
forcing $\sigma_{int} = 0$ and setting to $1$ the denominator in
the second term. As discussed in D'Agostini (2005), this is likely
to bias low the estimate of $\sigma_{int}$ which is just what we
find. Note that the scatter of the data around the best fit line
is essentially the same as the intrinsic scatter so that it is
fully explained by the sources generating $\sigma_{int}$, namely
the variation in the halo parameters.

\begin{figure}
\includegraphics[width=8.5cm]{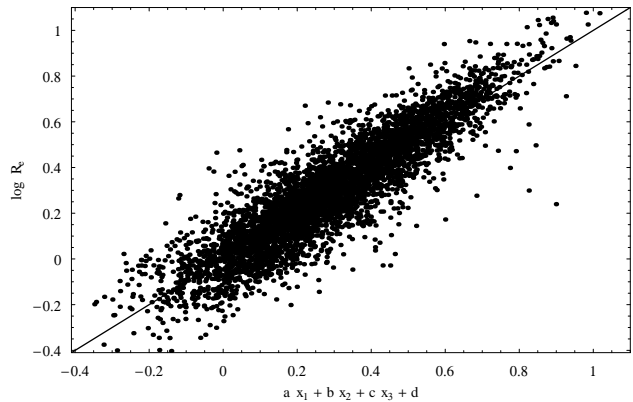}
\caption{Best fit curve (according to the Bayesian method)
superimposed on the $i'$ filter data for the fit in Eq.(\ref{eq:
svhlogre}).} \label{fig: svhlogre}
\end{figure}

Looking at the Table 1, it is not possible to definitively infer
whether the SVH coefficients are filter dependent. On the one
hand, the estimates for the different wavelengths are formally not
consistent within each other. Nevertheless, the differences are so
small that it is hard to say whether they are significant or due
to selection effects or some other uncontrolled systematics.

Being an immediate consequence of the virial theorem and our
assumptions (\ref{eq: linearratio}) and (\ref{eq: linearups}), we
have up to now written the SVH as a relation giving the velocity
dispersion as a function of the photometric parameters. However,
we can solve Eq.(\ref{eq: svh}) with respect to $R_e$ as\,:

\begin{equation}
\log{R_e} = a_{e} \log{\sigma_0} + b_e \log{\langle I_e \rangle} +
c_e \log{(n/4)} + d \label{eq: svhlogre}
\end{equation}
with\,:

\begin{equation}
\left \{
\begin{array}{l}
a_e = 1/b \\
b_e = -a/b \\
c_e = -c/b \\
d_e = -d/b \\
\end{array}
\right .  \label{eq: svhparlogre}
\end{equation}
which can be directly estimated for the different filters from the
values in Table 1. A most reliable estimate is, however, obtained
by fitting Eq.(\ref{eq: svhlogre}) to the data using the same
procedure as above thus obtaining also a value for the intrinsic
scatter. The results are reported in Table 2, while
Fig.\,\ref{fig: svhlogre} shows the best fit line superimposed to
the data in the fiducial $i'$ filter. It is immediate to see that
the values in Table 2 are in strong disagreement with the
theoretical expectations based on Eq.(\ref{eq: svhparlogre}) and
the values in Table 1. This is however by no means a shortcoming
of the model, but only an expected consequence of problems with
fitting an inverse relation so that we do not discuss anymore this
issue.

It is more interesting to note that the scatter in $\log{R_e}$ is
the same as the intrinsic one. Although this could also be
somewhat a consequence of neglecting uncertainties on $\log{R_e}$,
it is an intriguing result. Indeed, while in the FP the scatter
must be accounted for by both observational errors and intrinsic
terms, here the intrinsic scatter is the only source of scatter.
Moreover, such an intrinsic scatter is fully related to the
variations in the dark matter parameters so that its origin is
fully known and (to some extent) predictable provided a reliable
$c - M_{vir}$ relation is given.

It is worth wondering how the inverse SVH (\ref{eq: svhlogre})
works as distance indicator. To this aim, we note that the scatter
in $\log{R_e}$ translates into a $\ln{10} \times \sigma_{int,e}
\simeq 23\%$ scatter in the distance estimates. Table 4 in
Bernardi et al. (2003) reports a collection from literature of
various FP determinations with the corresponding scatter on the
distance ranging from $13\%$ to $22\%$ so that, taken at face
values, increasing the number of parameters with respect to the FP
(4 instead of 3 for the SVH vs the FP) does not ameliorate the
performances as distance indicator. Some remarks are however in
order. First, different determinations rely on different fitting
methods. Should we have used the direct fit approach, the scatter
in the distance should be reduced to $\sigma \simeq 16\%$ which is
smaller than the $20\%$ value obtained by Bernardi et al. (2003)
for a SDSS sample comprising 9000 ETGs with $0.01 \le z \le 0.3$.
A meaningful comparison should therefore rely on the same fitting
method and we advocate the use of our Bayesian approach in order
to not underestimate the intrinsic scatter. As a second issue, one
has also to consider that our sample is made out of all ETGs in
the lowZ catalog without any separation in field or clusters
objects. It is therefore worth reconsidering the fit and hence the
scatter by performing the fit on smaller subsamples separated
according to their environment. This is outside the aims of this
introductory investigation and will be presented elsewhere.

\section{Discussion and conclusions}

Early type galaxies may be considered as a homogenous class of
objects from many point of views. A further support to this idea
is represented by the existence of several interesting scaling
relations among their photometric and/or kinematic parameters, the
most famous ones being the FP (between $\log{\sigma_0}$,
$\log{\langle I_e \rangle}$ and $\log{R_e}$) and the PhP (where
$\log{\sigma_0}$ is replaced by the Sersic index $n$). In an
attempt to look for a unified description of both these relations,
we have presented here the {\it Sersic Virial Hyperplane} (SVH)
expressing a kinematical quantity, namely the velocity dispersion
$\sigma_0$ as a function of the Sersic photometric parameters $(n,
, R_e, \langle I_e \rangle)$. In the usual logarithmic units, such
a relation reduces to a hyperplane (i.e., a plane in four
dimensions) where all ETG lay with a small thickness as inferred
from the low intrinsic scatter $\sigma_{int}$. Just as the FJ
relation is a projection of the FP and the KR a projection of the
PhP, thus we find that both the FP and the PhP are projections of
the SVH so that the scatter in these well known relations can be
ascribed to neglect one of the SVH variables.

Our derivation of the SVH relies on very few assumptions. First,
ETGs are postulated to be in dynamical equilibrium so that the
virial theorem applies. Given that ETGs are old systems likely to
have formed most of their stellar content and settled their main
structural properties at $z \sim 2$ \cite{MC06}, this hypothesis
seems to be well founded, at least as a first well motivated
approximation. Starting from this premise, the SVH comes out as a
consequence of the virial theorem and of Eqs.(\ref{eq:
linearratio}) and (\ref{eq: linearups}). The first one relies on
an approximated log\,-\,linear relation between the model
dependent quantity $w({\bf p})/k({\bf p})$ and the Sersic
photometric parameters, while the second one assumes the existence
of a power\,-\,law relation between the stellar $M/L$ ratio
$\Upsilon_{\star}$ and the total luminosity $L_T$. Given mass
models for the luminous and dark components of a typical ETG,
Eq.(\ref{eq: linearratio}) has been verified by realistic
simulations reproducing the distribution in the photometric
parameters of a large ETG sample selected from the low redshift
version of the VAGC catalog based on SDSS DR2. On the other hand,
the relation $\Upsilon_{\star} \propto L_T^{\beta}$ is expected on
the basis of stellar populations synthesis models and is also
invoked to explain the FP tilt. It is worth stressing, however,
that here the tilt of the SVH is due to the dependence of the
stellar $M/L$ on $L_T$, while, in the case of the FP, such a
relation involve the global (stellar plus dark) $M/L$ ratio. As a
consequence, one has to resort to a mechanism coupling the dark
and luminous mass density profiles in order to have $M_{dyn}/L_T
\propto L_T^{\beta}$, while here we only rely on what is predicted
by stellar population synthesis models. From this point of view,
therefore, the SVH relies only on known physics without the need
of any unexplained interaction between baryons and CDM particles.

The discussion in Sect. 5 have confidently demonstrated that the
observed SVH is consistent with our theoretical predictions so
that it is worth wondering how the present introductory work can
be ameliorated. To this aim, it is worth reconsidering the
derivation of the observed SVH in Sect. 6. Here we have used the
Bayesian approach to infer the estimate of the SVH coefficients
and its intrinsic scatter from a dataset based on the lowZ
catalog. Such a sample is however affected by known problems.
First, the automated code used by Blanton et al. (2005) to
estimate the Sersic parameters does not work very well. The
recovered values of $(n, R_e, A)$ and hence of $\log{\langle I_e
\rangle}$ are biased in a complicated way depending on the values
of the parameters themselves. Modeling this bias and taking care
of it in the fitting procedure is quite difficult, but the problem
should be coped with by checking how it affects the estimate of
the SVH coefficients. It is worth stressing, however, that this
problem does not impact anyway our testing of the SVH. However, to
test the SVH, one has to verify that, for a given distribution of
photometric parameters based on a whichever real sample, the
simulated sample and the observed one predict the same SVH
coefficients which is indeed what we find. Note also that this
method allows to automatically include the selection effects in
the simulation without the need for accurately modelling them. The
data we have used present, however, a second shortcoming, i.e.
there is no estimate of the uncertainties on the Sersic
photometric parameters. As shown in Eq.(\ref{eq: lnl}), these
quantities enter the definition of the merit function to be
minimized in the Bayesian determination of the SVH coefficients.
Although one can estimate what the effect is by artificially
attaching measurement errors to the input quantities, it is more
convenient to solve the problem by resorting to a different sample
in order to address the problem hinted above. The recently
released Millenium Galaxy Catalog \cite{MGC03,MGC06} contains a
detailed bulge/disc decomposition of $\sim 10000$ nearby galaxies
with detailed fitting of the Sersic law to the surface brightness
profile \cite{A06}. The code is tested and shown not to be biased
and the errors on the photometric parameters are available.
Cross\,-\,matching with the SDSS and selecting only the ETGs
should provide an ideal sample to test the SVH retrieving a more
reliable estimate of its coefficients and scatter thus allowing to
reconsider it as a distance indicator.

From a theoretical point of view, it is worth reconsidering our
basic assumptions. As discussed above, the tilt of the SVH with
respect to the virial theorem predictions may be fully ascribed to
a power\,-\,law relation between the stellar $M/L$ ratio
$\Upsilon_{\star}$ and the total luminosity $L_T$. Although
estimating the slope $\beta$ from the tilt is, in principle,
possible, confronting with an expected value is welcome. To this
aim, one can resort to stellar population synthesis models
\cite{FRV97,BG03,LB05,Mar05} by varying the different ingredients
entering the codes and looking (by trial and error) for the
combination giving the slope $\beta$ needed to reproduce the
correct SVH tilt. Should these stellar models be able to reproduce
the observed colors of ETG, our derivation of the SVH could be
further strengthened.

A fundamental role in the ETG modeling has been played by the
choice of the dark halo model. Although the NFW mass density
profile is the standard one, it is nevertheless well known that it
encounters serious difficulties in explaining the inner rotation
curves of low surface brightness galaxies (see, e.g., de Blok 2005
and refs. therein). Moreover, some recent evidences from the
planetary nebulae dynamics have put into question the need for a
significative amount of dark matter in the inner regions of
elliptical galaxies \cite{Rom03,Nic05}. Estimating the dark matter
content for the real galaxies is difficult, but we can rely on
what we know from the modeling of the simulated galaxies since
they reproduce the same SVH as the observed ones. Defining the
dynamical mass\footnote{Note that here we compute the dynamical
mass from the known values of the model parameters rather than
estimating it from the velocity dispersion. For this reason,
although the definition is the same, our values for $M_{dyn}$
differ from those one would obtain using the formula in
Padmanabhan et al. (2005).} $M_{dyn}$ as the total mass within
$R$, i.e. $M_{dyn}(R) = \Upsilon_{\star} L(R) + M_{DM}(R)$, we
find a median value $M_{DM}(R_e)/M_{dyn}(R_e) \simeq 14\%$ (with a
rms value of $20\%$) so that inner regions turn out to be baryon
dominated as expected. This also makes us confident that changing
the halo model does not affect significantly the theoretical
values of the SVH coefficients. A subtle effect, worth to be
investigated, is the relation between the concentration $c$ and
the virial mass $M_{vir}$ of the halo. Actually, the scatter of
this relation concurs in determining the scatter in the SVH. Said
in another way, for given $M_{vir}$ and stellar mass parameters,
the scatter in the $c\,-\,M_{vir}$ relation introduces a scatter
on $M_{DM}(R_e)/M_{dyn}(R_e)$ thus contributing to the total SVH
scatter. Numerical N\,-\,body simulations and semi\,-\,analytical
galaxy formation models predict different halo models with its own
$c\,-\,M_{vir}$ relation and scatter. It should be tempting to
investigate whether a large ETG sample (free of the problems
described above) could be used to discriminate among these
different possibilities on the basis of the scatter they predict
for the SVH.

The aim of the present paper was mainly to introduce the SVH as a
unifying scenario for the ETG scaling relations. If the
encouraging results presented here will be further confirmed, both
observationally and theoretically, we are confident that the SVH
could represent a valid tool to investigate the ETG properties
under a single picture. \\

{\it Acknowledgements.} Funding for the Sloan Digital Sky Survey
(SDSS) has been provided by the Alfred P. Sloan Foundation, the
Participating Institutions, the National Aeronautics and Space
Administration, the National Science Foundation, the U.S.
Department of Energy, the Japanese Monbukagakusho, and the Max
Planck Society. The SDSS Web site is http://www.sdss.org/.

The SDSS is managed by the Astrophysical Research Consortium (ARC)
for the Participating Institutions. The Participating Institutions
are The University of Chicago, Fermilab, the Institute for
Advanced Study, the Japan Participation Group, The Johns Hopkins
University, Los Alamos National Laboratory, the
Max-Planck-Institute for Astronomy (MPIA), the
Max-Planck-Institute for Astrophysics (MPA), New Mexico State
University, University of Pittsburgh, Princeton University, the
United States Naval Observatory, and the University of Washington.


\begin{thebibliography}{99}

\bibitem[\protect\citename{Abazajian et al. }2004]{DR2}
Abazajian, K., Adelman\,-\,McCarthy, J.K., Ag\"ueros, M.A., Allam,
S.S., Anderson, K.S.J., Anderson, S.F. et al. 2004, AJ, 128, 502

\bibitem[\protect\citename{Adelman\,-\,McCarthy et al. }2006]{DR4}
Adelman\,-\,McCarthy, J.K., Ag\"ueros, M.A., Allam, S.S.,
Anderson, K.S.J., Anderson, S.F. et al. 2006, ApJS, 162, 38

\bibitem[\protect\citename{Akritas \& Bershady }1996]{AB96}
Akritas, M.G., Bershady, M.A. 1996, ApJ, 470, 706

\bibitem[\protect\citename{Allen et al. }2006]{A06}
Allen, P.D., Driver, S.P., Graham, A.W., Cameron, E., Liske, J.,
de Propris, R. 2006, MNRAS, 371, 2

\bibitem[\protect\citename{Bender et al. }1992]{BBF92}
Bender, R., Burstein, D., Faber, S.M. 1992, ApJ, 399, 462

\bibitem[\protect\citename{Bernardi et al. }2003]{BernFP}
Bernardi, M., Sheth, R.K., Annis, J., Burles, S., Eisenstein, D.J.
et al. 2003, AJ, 125, 1866

\bibitem[\protect\citename{Bernardi et al. }2005]{B05}
Bernardi, M., Sheth, R.K., Nichol, R.C., Schneider, D.P.,
Brinkmann, J. 2005, AJ, 129, 61

\bibitem[\protect\citename{Binney \& Tremaine }1987]{BT87}
Binney, J., Tremaine, S. 1987, {\it Galactic Dynamics},  Princeton
University Press, Princeton

\bibitem[\protect\citename{Blanton et al. }2005]{VAGC}
Blanton, M.R., Schlegel, D.J., Strauss, M.A., Brinkmann, J.,
Finkbeiner, D. et al. 2005, AJ, 129, 2562

\bibitem[\protect\citename{Bruzual \& Charlot }2003]{BG03}
Bruzual, G., Charlot, S. 2003, MNRAS, 344, 1000

\bibitem[\protect\citename{Bullock et al. }2001]{Buletal01}
Bullock, J.S., Kolatt, T.S., Sigad, Y., Somerville, R.S.,
Kravtsov, A.V., Klypin, A., Primack, J.P., Dekel, A. 2001, MNRAS,
321, 559

\bibitem[\protect\citename{Busarello et al. }1997]{Bus97}
Busarello. G., Capaccioli, M., Capozziello, S., Longo, G., Puddu,
E. 1997, A\&A, 320, 415

\bibitem[\protect\citename{Caon et al. }1993]{CCD93}
Caon, N., Capaccioli, M., D' Onofrio, M. 1993, MNRAS, 265, 1013

\bibitem[\protect\citename{Cardone et al. }2005]{PoLLS}
Cardone, V.F., Piedipalumbo, E., Tortora, C. 2005, MNRAS, 358,
1325

\bibitem[\protect\citename{Ciotti }1991]{C91}
Ciotti, L. 1991, A\&A, 249, 99

\bibitem[\protect\citename{D'Agostini }2004]{Dago04}
D'Agostini, G. 2004, physics/0412148

\bibitem[\protect\citename{D'Agostini }2005]{Dago05}
D'Agostini, G. 2005, physics/0511182

\bibitem[\protect\citename{Dalal \& Keeton }2003]{DK03}
Dalal, N., Keeton, C.R. 2003, astro\,-\,ph/0312072

\bibitem[\protect\citename{de Blok }2005]{db05}
de Blok, W.J.G. 2005, ApJ, 634, 227

\bibitem[\protect\citename{de Vaucouleurs }1948]{deV48}
de Vaucouleurs, G. 1948, Ann. d' Astroph., 11, 247

\bibitem[\protect\citename{Djorgovski \& Davis }1987]{DD87}
Djorgovski, S., Davis, M. 1987, ApJ, 313, 59

\bibitem[\protect\citename{Dressler et al. }1987]{7S87}
Dressler, A., Lynden\,-\,Bell, D., Burstein, D., Davies, R.L.,
Faber, S.M., Terlevich, R.J., Wegner, G. 1987, ApJ, 313, 42

\bibitem[\protect\citename{Driver et al. }2006]{MGC06}
Driver, S.P., Allen, P.D., Graham, A.W., Cameron, E., Liske, J.
2006, MNRAS, 368, 414

\bibitem[\protect\citename{Faber \& Jackson }1976]{FJ76}
Faber, S., Jackson, R. 1976, ApJ, 317, 1

\bibitem[\protect\citename{Feigelson \& Babu }1996]{FB92}
Feigelson, E.D., Babu, G.J. 1992, ApJ, 397, 55

\bibitem[\protect\citename{Fioc \& Rocca\,-\,Volmerange }1996]{FRV97}
Fioc, M., Rocca\,-\,Volmerange, B. 1997, A\&A, 326, 950

\bibitem[\protect\citename{Fukugita et al. }1995]{FSI95}
Fukugita, M., Shimasaku, K., Ichikawa, T. 1995, PASP, 107, 945

\bibitem[\protect\citename{Fukugita et al. }1998]{FHP98}
Fukugita, M., Hogan, C.J., Peebles, P.J.E. 1998, ApJ, 503, 518

\bibitem[\protect\citename{Ghigna et al. }2000]{Ghigna00}
Ghigna, S., Moore, B., Governato, F., Lake, G., Quinn, T., Stadel,
J. 2000, ApJ, 544, 616

\bibitem[\protect\citename{Graham \& Colless }1997]{GC97}
Graham, A.W., Colless, M. 1997, MNRAS, 287, 221

\bibitem[\protect\citename{Graham }2002]{G02}
Graham, A.W. 2002, MNRAS, 334, 859

\bibitem[\protect\citename{Graham \& Driver }2005]{GD05}
Graham, A.W., Driver, S.P. 2005, PASA, 22, 118

\bibitem[\protect\citename{Graham et al. }2006a]{G06a}
Graham, A.W., Merritt, D., Moore, B., Diemand, J., Terzic, B.
2006a, AJ, 132, 2701

\bibitem[\protect\citename{Graham et al. }2006b]{G06b}
Graham, A.W., Merritt, D., Moore, B., Diemand, J., Terzic, B.
2006b, AJ, 132, 2711

\bibitem[\protect\citename{J$\o$rgensen et al. }1995]{JFK95}
J$\o$rgensen, I., Franx, M., Kj$\ae$rgaard, P. 1995, MNRAS, 273,
1097

\bibitem[\protect\citename{J$\o$rgensen et al. }1995]{JFK96}
J$\o$rgensen, I., Franx, M., Kj$\ae$rgaard, P. 1996, MNRAS, 280,
167

\bibitem[\protect\citename{Kauffman et al. }2003]{K03}
Kauffman, G., Heckman, T.M., White, S.D.M., Charlot, S., Tremonti,
C. et al. 2003, MNRAS, 341, 33

\bibitem[\protect\citename{Khosroshahi et al. }2000]{KWKM00}
Khosroshahi, H.G., Wadadekar, Y., Kembhavi, A., Mobasher, B. 2000,
ApJ, 531, L103

\bibitem[\protect\citename{Kormendy }1977]{K77}
Kormendy, J. 1977, ApJ, 218, 333

\bibitem[\protect\citename{La Barbera et al. }2000]{LaB00}
La Barbera, F., Busarello, G., Capaccioli, M. 2000, A\&A, 362, 851

\bibitem[\protect\citename{La Barbera et al. }2004]{LaB04}
La Barbera, F., Merluzzi, P., Busarello, G., Massarotti, M.,
Mercurio, A. 2004, A\&A, 425, 797

\bibitem[\protect\citename{La Barbera et al. }2005]{LaB05}
La Barbera, F., Covone, G., Busarello, G., Capaccioli, M., Haines,
C.P., Mercurio, A., Merluzzi, P. 2005, MNRAS, 358, 1116

\bibitem[\protect\citename{Le Borgne et al. }2005]{LB05}
Le Borgne, D., Rocca\,-\,Volmerange, B., Prugniel, P., Lancon, A.,
Fioc, M., Soubiran, C. 2005, A\&A, 425, 881

\bibitem[\protect\citename{Lima Neto et al. }1999]{LGM99}
Lima Neto, G.B., Gerbal, D., M\'arquez, I. 1999, MNRAS, 309, 481

\bibitem[\protect\citename{Liske et al. }2003]{MGC03}
Liske, J., Lemon, D.J., Driver, S.P., Cross, N.J.G., Couch, W.J.
2003, MNRAS, 344, 307

\bibitem[\protect\citename{Maraston }2005]{Mar05}
Maraston, C. 2005, MNRAS, 362, 799

\bibitem[\protect\citename{M\'arquez et al. }2001]{Metal01}
M\'arquez, I., Lima Neto, G.B., Capelato, H., Durret, F., Lanzoni,
B., Gerbal, D. 2001, A\&A, 379, 767

\bibitem[\protect\citename{Mazure \& Capelato }2002]{MC02}
Mazure, A., Capelato, H.V. 2002, A\&A, 383, 384

\bibitem[\protect\citename{Merlin \& Chiosi }2006]{MC06}
Merlin, E., Chiosi, C. 2006, A\&A, 457, 437

\bibitem[\protect\citename{Merritt et al. }2005]{Merr05}
Merritt, D., Graham, A.W., Moore, B., Diemand, J., Terzic, B.
2005, astro\,-\,ph/0509417 (AJ in press)

\bibitem[\protect\citename{Moore et al. }1998]{Moore98}
Moore, B., Governato, F., Quinn, T., Stadel, J. 1998, ApJ, 499, L5

\bibitem[\protect\citename{Moore et al. }1999]{moo+al99}
Moore, B., Ghigna, S., Governato, F., Lake, G., Quinn, T., Stadel,
J., Tozzi, P., 1999, ApJ, 524, L19

\bibitem[\protect\citename{Napolitano et al. }2005]{Nic05}
Napolitano, N.R., Capaccioli, M., Romanowsky, A.J., Douglas, N.G.,
Merrifield, M.R., Kuijken, K., Arnaboldi, M., Gerhard, O.,
Freeman, K.C. 2005, MNRAS, 357, 691

\bibitem[\protect\citename{Navarro et al. }1997]{NFW97}
Navarro, J.F., Frenk, C.S., White, S.D.M. 1997, ApJ, 490, 493

\bibitem[\protect\citename{Navarro et al. }2004]{N04}
Navarro, J.F., Hayashi, E., Power, C., Jenkins, A.R., Frenk, C.S.
et al. 2004, MNRAS, 349, 1039

\bibitem[\protect\citename{Noordermeer et al. }2007]{N07}
Noordermeer, E., van der Hulst, J.M., Sancisi, R., Swaters, R.S.,
van Albada, T.S. 2007, astro\,-\,ph/0701731

\bibitem[\protect\citename{Padmanabhan et al. }2004]{Pad04}
Padmanabhan, N., Seljak, U., Strauss, M.A., Blanton, M.R.,
Kauffman, G. et al. 2004, New Astron., 9, 329

\bibitem[\protect\citename{Prugniel \& Simien }1997]{PS97}
Prugniel, Ph., Simien, F. 1997, A\&A, 321, 111

\bibitem[\protect\citename{Roscoe }1999]{R99}
Roscoe, D.F. 1999, A\&A, 343, 788

\bibitem[\protect\citename{Romanowski et al. }2003]{Rom03}
Romanowsky, A.J., Douglas, N.G., Arnaboldi, M., Kuijken, K.,
Merrifield, M.R., Napolitano, N.R.; Capaccioli, M., Freeman, K.C.
2003, Science, 301, 1696

\bibitem[\protect\citename{Sersic }1968]{Sersic}
Sersic, J.L. 1968, {\it Atlas de Galaxies Australes}, Observatorio
Astronomico de Cordoba

\bibitem[\protect\citename{Shimasaku et al. }2001]{Shima01}
Shimasaku, K., Fukugita, M., Doi, M., Hamabe, M., Ichikawa, T. et
al., 2001, AJ 122, 1238

\bibitem[\protect\citename{Trujillo et al. }2004]{TBB04}
Trujillo, I., Bukert, A., Bell, E.F. 2004, ApJ, 600, L39

\bibitem[\protect\citename{van der Marel }1991]{vdM91}
van der Marel, R. 1991, MNRAS, 253, 710

\bibitem[\protect\citename{Young \& Currie }1994]{YC94}
Young, C.K., Currie, M.J. 1994, MNRAS, 268, L11

\end{thebibliography}
\end{document}